\documentclass[aps,prd,preprint,showpacs,amsmath,nofootinbib,amssymb,eps]{revtex4}
\usepackage{graphicx,amsmath,color}
\usepackage{epsfig}
\usepackage{dcolumn}
\usepackage{bm}

\newcommand{\beq} {\begin{equation}}
\newcommand{\eeq} {\end{equation}    }
\newcommand{\bea} {\begin{eqnarray} }
\newcommand{\eea} {\end{eqnarray}    }
\newcommand{\fbar}{\bar{f}}

\newcommand{\im}{{\Im m  }}

\newcommand{\Lg}{{\mathcal L}}
\newcommand{\idthree}{{\mathbb I \/}_3}

\newcommand{\lm}{\lambda}

\newcommand{\no}{\nonumber}
\newcommand{\gm}{\gamma}

\newcommand{\Gm}{\Gamma}

\newcommand{\sq}{\sqrt{2}}

\newcommand{\dt}{\delta}

\newcommand{\rd}{\partial}

\newcommand{\br}{{\rm Br}}

\newcommand{\gev}{~{\rm GeV}}
\newcommand{\tev}{~{\rm TeV}}

\newcommand{\cw}{c_W}
\newcommand{\tw}{t_W}
\newcommand{\sbt}{s_{\beta}}

\newcommand{\cbt}{c_{\beta}}
\newcommand{\tbt}{t_\beta}

\newcommand{\xlm}{x_{\lambda}}
\newcommand{\shat}{{\hat{s}}}

\newcommand{\lsim}{\mathrel{\mathop{\kern 0pt \rlap
  {\raise.2ex\hbox{$<$}}}
  \lower.9ex\hbox{\kern-.190em $\sim$}}}
\newcommand{\gsim}{\mathrel{\mathop{\kern 0pt \rlap
  {\raise.2ex\hbox{$>$}}}
  \lower.9ex\hbox{\kern-.190em $\sim$}}}

\begin{document}
\title {Production and decays of the light pseudoscalar boson $\eta$
at the LHC\\
in the simplest little Higgs model}
\author{
Kingman Cheung$^{1,2}$, Jeonghyeon Song$^3$, Poyan Tseng$^1$, Qi-Shu Yan$^{1,2}$}
\affiliation{
$^1$ Department of Physics, National Tsing Hua University,
Hsinchu, Taiwan \\
$^2$
Physics Division, National Center for Theoretical Sciences, Hsinchu, Taiwan\\
$^3$ Department of Physics, Konkuk University, Seoul 143-701, Korea
}
\renewcommand{\thefootnote}{\arabic{footnote}}
\date{\today}

\begin{abstract}
In many extensions of the standard model, the Higgs sector often
contains an additional pseudoscalar boson. A good example is the
SU(3) simplest little Higgs model, which accommodates a light
pseudoscalar boson $\eta$ with quite different characteristics from
those in other multi-Higgs-doublet models. We study various phenomenological signatures
of the $\eta$ at the LHC. 
In particular, we calculate in details both production and decays in
the Drell-Yan type channel $q \bar{q} \to Z/Z' \to h \eta$, and in
the associated production with a $t\bar t$ pair, $gg\,(q\bar q) \to t \bar t \eta$.
We emphasize the $\tau^+ \tau^-$ decay mode of the $\eta$ boson when its
mass is below the $b\bar b$ threshold.
We show that $t\bar t \eta$ production is in fact large enough to give a sizable
number of events while suppressing the backgrounds.
We also comment on the direct gluon fusion process and the indirect decay
from the heavy $T$ quark ($T \to t \eta$).
\end{abstract}

\maketitle


\section{Introduction}
\label{sec:Introduction}

One of the primary goals of the CERN Large Hadron Collider (LHC) is
to understand the electroweak symmetry breaking, which is vital in
explaining the origin of the fermion and gauge boson masses. In the
standard model (SM), one single Higgs doublet is introduced to
trigger the electroweak symmetry breaking, of which the by-product
is a scalar boson known as the Higgs boson.
Although the SM
does not predict  the mass of the Higgs boson ($m_H$), we have direct, indirect,
and theoretical bounds on $m_H$.
The most up-to-date search has placed a lower bound of 114.4
GeV\;\cite{LEP} on $m_H$.
The precision measurements from LEP and SLD collaborations
have placed an upper bound on $m_H < 160$ GeV
(at one-sided 95\% C.L.)\;\cite{prec}, which is much lighter than the
theoretical one (the so-called triviality bound) of about 1 TeV.
The consensus is that the Higgs boson is rather light.

The SM by itself cannot provide any theoretical framework
to guarantee the lightness of the Higgs boson.
Very often small masses are protected by some symmetries, \textit{e.g}.,
the chiral symmetry to protect fermion masses and
the gauge symmetry to protect gauge boson masses. There are no such
symmetries in the SM to protect the scalar boson masses.
A recent class of models, dubbed the little Higgs models, has been developed
based on the idea that the lightness of the Higgs boson is attributed to
its being a pseudo Nambu-Goldstone boson (pNGB)\;\cite{LH}.
Armed by the collective symmetry breaking idea,
little Higgs models can explain the little hierarchy problem.
The Higgs boson mass is radiatively generated
with quadratic divergence emerging at two
loop level:
The Higgs boson mass around 100 GeV and the 10 TeV cut-off
are possible without fine-tuning.
The one-loop level quadratic divergences
from the SM gauge boson and top-quark loops are
canceled by those from new heavy gauge boson and heavy $T$-quark loops,
respectively.
According to the global symmetry breaking pattern
little Higgs models can be classified into two categories: (i)
the \textit{product group} models where the diagonal breaking of
two (or more) gauge groups leads to the SM gauge group, and (ii)
the \textit{simple group} models where a single
larger gauge group is broken into the SM gauge group.
The most studied \textit{product group} model is the littlest Higgs
model\;\cite{littlest}
while that for the \textit{simple group} model is
the simplest little Higgs model\;\cite{simplest}.  In this work, we
focus on the simplest little Higgs (SLH) model, which generates the least
fine tuning in explaining the low Higgs mass\;\cite{fine-tuning}.

A special feature of the simplest little Higgs model is the presence
of a light pseudoscalar boson, denoted by $\eta$.
The model is based on [SU(3) $\times$ U(1)$_X]^2$ global
symmetry with its diagonal subgroup SU(3) $\times$ U(1)$_X$ gauged.
The vacuum expectation values (VEV) of two SU(3)-triplet scalar
fields, $\langle\Phi_{1,2}\rangle = (0,0,f_{1,2})^T$, spontaneously
break both the global symmetry and the gauge symmetry.  Uneaten pNGB's
consist of a SU(2)$_L$ doublet $h$ and a pseudoscalar $\eta$.
In Ref.\;\cite{cheung-song} it was pointed out that the $\eta$ boson in the original model
is massless, which is problematic for $\eta$
production in rare $K$ and $B$ decays, $B$-$\bar{B}$ mixing, and
$\Upsilon \to \eta\gamma$, as well as for the cosmological axion
limit.
One of the simplest remedies was suggested
by introducing a $-\mu^2(\Phi_1 ^\dagger \Phi_2+ h.c.)$
term into the scalar potential by hand\;
\cite{simplest,Kaplan:Schmaltz,Kilian:pseudo-scalar}.
This $\mu$ term then determines the $\eta$ mass.
The mass of $\eta$ is not theoretically constrained, but there
exists an experimental constraint from non-observation in the decay
$\Upsilon \to \gamma + X_0$.  It excludes pseudoscalar bosons with
mass below 5--7 GeV\;\cite{heavyQ}.
It has been also shown
that a sizable portion of parameter space
kinematically allows the decay $h \to \eta \eta$,
which can relieve the constraint on the direct search bound on the Higgs
boson mass\;\cite{cheung-song,cheung-song-yan}.

In this work, we focus on production and decays of a light $\eta$ boson at
the LHC.  The decay pattern of $\eta$ is quite similar to that of the
SM Higgs boson.  A few distinctive features are (i) the $\eta$ does not
decay into $WW$ and $ZZ$, (ii) $\eta$ has a rather large
branching ratio into $g g$, and (iii) the dominant decay mode is
$\eta \to Z H$ if kinematically allowed.
The largest production channel for $\eta$ is gluon fusion, but the
decay of $\eta \to b\bar b,jj$ will be buried under QCD backgrounds
while $\eta \to \tau^+ \tau^-$ will not likely stand out of the
Drell-Yan background.  The $WW$ fusion does not contribute to $\eta$
production.  Associated production $\eta$ with $t\bar t$ pair and with
the Higgs boson could be the most useful channels to search for
the $\eta$.

The organization is as follows.  In the next section, we describe briefly
the simplest little Higgs model with the $\mu$ term.  We calculate
the decays and production of the $\eta$ boson in Sec.\,III and IV,
respectively.  We study the detection of the $\eta$ boson in Sec. V and VI.
We then conclude in Sec.\,VII.

\section{$SU(3)$ simplest group model with the $\mu$ term}
\label{sec:SLH}
The SU(3) simplest little Higgs model is based on
$[\,\mathrm{SU}(3) \times \mathrm{U}(1)_X]^2$ global symmetry
with its diagonal subgroup $\mathrm{SU}(3) \times \mathrm{U}(1)_X$ gauged.
The symmetry breaking of
$\mathrm{SU}(3) \times \mathrm{U}(1)_X \to \mathrm{SU}(2)_L \times \mathrm{U}(1)_Y$
is generated by aligned VEVs of two complex SU(3)
triplet scalar fields,
$\Phi_1$ and $\Phi_2$.
Out of the 10 degrees of freedom in $\Phi_1$ and $\Phi_2$,
five are eaten by the SU(3) symmetry breaking.
Remained five degrees of freedom in $\Phi_{1,2}$
are parameterized as a nonlinear sigma model with
\beq
    \Phi_1 = e^{i t_\beta \Theta} \Phi_{1}^{(0)}
        , \quad
    \Phi_2 = e^{-i \Theta/t_\beta } \Phi_{2}^{(0)}
        ,
\eeq
where $t_\beta\equiv\tan\beta $ and
\begin{equation}
   \Theta = \frac{1}{f} \left[
        \left( \begin{array}{cc}
        \begin{array}{cc} 0 & 0 \\ 0 & 0 \end{array}
            & h \\
        h^{\dagger} & 0 \end{array} \right)
        + \frac{\eta}{\sqrt{2}}
        \left( \begin{array}{ccr}
        1 & 0 & 0 \\
        0 & 1 & 0 \\
        0 & 0 & 1 \end{array} \right) \right]
        \equiv \frac{h_0}{f} \,\widehat{\mathbb{H}}  +\frac{\eta}{\sq f} \idthree .
\end{equation}

Radiatively generated VEV of the Higgs boson field $h$
triggers the SM electroweak symmetry breaking (EWSB):
The Higgs boson is defined by $h = (v + h_0 )/\sqrt{2}$.
Without resort to any expansion, $\Phi_1$ and $\Phi_2$ have
the following closed form\;\cite{cheung-song}:
\beq
\label{eq:Phi12}
    \Phi_1 = 
f \cbt
e^{ i\frac{\tbt\eta}{\sq f}}
\left(
  \begin{array}{c}
    i \sin  \frac{\tbt h_0}{f}\\
    0 \\
    \cos \frac{\tbt h_0}{f}  \\
  \end{array}
\right)
                                   ,
                                   \quad
    \Phi_2 =
                                  f \sbt
e^{ -i\frac{\eta}{\sq\tbt f}}
\left(
  \begin{array}{c}
    - i \sin \frac{h_0}{\tbt f} \\
    0 \\
    \cos \frac{h_0}{\tbt f} \\
  \end{array}
\right).
\eeq
Note that $\eta$ in $\Phi_{1,2}$ is only a phase factor.
Explicit form of $\partial_\mu \Phi_{1,2}$ are also
useful for later discussions:
\bea
\label{eq:DPhi12}
   \rd_\mu \Phi_1 &=&
   \phantom{-}
i \sbt e^{i \frac{\tbt \eta}{\sq f}}
\left\{
\frac{\rd \eta}{\sq}
+ \rd h_0 \widehat{\mathbb{H}}
\right\}\left(
  \begin{array}{c}
    i \sin  \frac{\tbt h_0}{f}\\
    0 \\
    \cos \frac{\tbt h_0}{f}  \\
  \end{array}
\right),
\\ \no
\rd_\mu    \Phi_2
&=& -i \cbt e^{-i \frac{ \eta}{\sq \tbt f}}
\left\{
\frac{\rd \eta}{\sq}
+ \rd h_0 \widehat{\mathbb{H}}
\right\}
\left(
  \begin{array}{c}
    - i \sin \frac{h_0}{\tbt f} \\
    0 \\
    \cos \frac{h_0}{\tbt f} \\
  \end{array}
\right).
\eea

The covariant derivative term is
\beq
\label{eq:Lg:gauge0}
\Lg_\Phi = \sum_{j=1,2}\left|
\left(\rd_\mu + i g A^a_\mu T^a - i \frac{g_x}{3} B^x_\mu \right)
\Phi_j
\right|^2
\equiv
\sum_{j=1,2}\left|
\left(\rd_\mu + i g \mathbb{G}_\mu \right)
\Phi_j
\right|^2,
\eeq
where $g_x =g\tw\,\sqrt{1-\tw^2/3}$ and $\tw$ is the tangent of the electroweak mixing angle.
Detailed expressions for $\mathbb{G}$ is referred to Ref.\;\cite{smoking}.
As $\Phi_1$ and $\Phi_2$ develop their VEVs,
the 5 degrees of freedom appear as the longitudinal component of the
heavy gauge bosons, including a $Z'$ gauge boson
and a complex SU(2) doublet $(Y^0,X^-)$,
with masses
\beq
M_{Z'}=\sqrt{\dfrac{2}{3-t_W^2}}\,g\, f, \quad M_{X^\pm}=M_Y=\dfrac{g f}{\sqrt{2}}
\,.
\eeq

For convenience we separate the Lagrangian in Eq.\,(\ref{eq:Lg:gauge0})
into three terms:
\bea
\label{eq:gauge:interaction}
\Lg_\Phi
&=&
\sum_{j=1,2}\left|
\rd_\mu
\Phi_j
\right|^2
+ \sum_{j=1,2} \left[ -i g \Phi_j^\dagger \mathbb{G}^\mu \rd_\mu \Phi_j + H.c. \right]
+ g^2 \sum_{j=1,2}\Phi_j^\dagger \mathbb{G}^\mu \mathbb{G}_\mu\Phi_j
\\ \no
&\equiv& \Lg_{kin} + \Lg_{int} + \Lg_{mass}.
\eea
Using Eq.\,(\ref{eq:Phi12})
it is easy to see that the first term
is just the kinetic term of the Higgs boson
and $\eta$:
\beq
\sum_{i=1,2}\left|\rd \Phi_i \right|^2 =\frac{(\rd \eta)^2}{2}  + (\rd h_0 )^2.
\eeq
The last term $\Lg_{mass}$
leads to the masses for the gauge bosons
as well as the coupling of the Higgs boson with two gauge boson.
Since $\eta$ is only a phase factor in $\Phi_{1,2}$,
the $\eta$-dependence in $\Lg_{mass}$ disappears.
{}From Eqs.\,(\ref{eq:Phi12})
and (\ref{eq:DPhi12}),
the second term $\Lg_{int}$ leads to
only the $\eta$-$H$-$Z$, $\eta$-$H$-$Z'$, and $\eta$-$H$-$\im(Y^0)$ couplings:
\bea
\label{eq:Z:h:eta}
\Lg_{int} &=& \sq
\left( \tbt - \frac{1}{\tbt} \right)
\frac{m_Z}{f}   (H \rd_\mu \eta - \eta \rd_\mu H)
\left[ Z^\mu
- f_{Z'} Z^{\prime\mu}
\right]
\\ \no &&
+  \sq  g (H \rd_\mu \eta - \eta \rd_\mu H)  \im (Y^{0 \mu}) ,
\eea
where $f_{Z'}= \cw (1-\tw^2)/\sqrt{3-\tw^2} \approx 0.682 $.

The fermion sector in this model should be extended
since the gauged SU(3) symmetry promotes
the SM fermions into SU(3) triplets.
The Yukawa interaction of the third generation quarks
is determined by the little Higgs mechanism
which cancels the largest contribution of the top quark
to the radiative Higgs mass $\dt m_H$.
However negligible contributions to $\dt m_H$ of the first two
generation quarks and all generation leptons
leave some ambiguity in fermion embedding.
In the literature, two kinds of fermion embedding
have been discussed, the
``universal'' embedding\;\cite{smoking}, and the ``anomaly-free''
embedding\;\cite{kong}.
In this paper we focus on the anomaly-free embedding case.
The universal embedding case has almost
the same
Yukawa couplings of $\eta$, except that
the first two generation heavy quarks are up-type
while those in the anomaly-free embedding are down-type.

The quark Yukawa interactions for the third generation and
for the first two generations are given by\,\cite{smoking}
\bea
L_3 &=&  i \lambda_1^t t_1^c \Phi_1^{\dagger} Q_3
+ i \lambda_2^t t_2^c \Phi_2^{\dagger} Q_3
+ i  \frac{\lambda_d^m}{\Lambda}  d_m^c \epsilon_{ijk} \Phi_1^i \Phi_2^j Q_3^k + H.c.\,, \\
L_{1,2} &=&  i \lambda_1^{d_n} d_{1n}^c Q_n^{T} \Phi_1 + i
\lambda_2^{d_n} d_{2n}^c Q^{T}_n \Phi_2 + i
\frac{\lambda_{u}^{mn}}{\Lambda} u_m^c \epsilon_{ijk} \Phi_1^{*i}
\Phi_2^{*j} Q_n^k + H.c.,
\eea
where  $n=1,2$;
$i,j,k=1,2,3$ are SU(3) indices;
$Q_3=\{ t_L,  b_L,  i T_L\}$ and
$Q_n = \{ d_{nL}, - u_{nL}, i D_{nL}\}$;
$d_m^c$ runs over $(d^c, s^c, b^c, D^c, S^c)$;
$u^c_m$ runs over $(u^c, c^c, t^c, T^c)$.

The mass eigenstate ($f^c$, $F^c$)
are the mixture of ($f_1^c$, $f_2^c$),
where $f=t,s,c$ and $F=T,S,C$,
\beq
\left(
  \begin{array}{c}
   t^c \\
    T^c \\
  \end{array}
\right)
=\left(
   \begin{array}{rr}
     - \cos \theta_T & \sin \theta_T \\
    \sin \theta_T & \cos \theta_T \\
   \end{array}
 \right)
 \left(
   \begin{array}{c}
      t_1^c  \\
     t_2^c  \\
   \end{array}
 \right),
 \quad
\left(
  \begin{array}{c}
   d^c  \\
    D^c \\
  \end{array}
\right)
=\left(
   \begin{array}{rr}
      \cos \theta_{D,S} & \sin \theta_{D,S} \\
    \sin \theta_{D,S} & -\cos \theta_{D,S} \\
   \end{array}
 \right)
 \left(
   \begin{array}{c}
      d_1^c  \\
     d_2^c  \\
   \end{array}
 \right),
\eeq
where the $s$ quark sector is the same as the $d$ quark sector.
The mixing angles are
\beq
\sin \theta_F =  \frac{\lambda^f_1 \cbt }{\sqrt{(\lambda^f_1 \cbt)^2
+ (\lambda^f_2 \sbt)^2 } }\,,\quad
\cos \theta_F = \frac{\lambda^f_2 \sbt}{\sqrt{(\lambda^f_1 \cbt)^2
+ (\lambda^f_2 \sbt)^2 } }\,,
\eeq
where $\cbt=\cos\beta$ and $\sbt=\sin\beta$.

The heavy quark masses ($M_T$, $M_S$, $M_D$)
and the SM quark masses are
\beq
M_Q = \sqrt{ (\lm_1^q \cbt)^2+(\lm_2^q \sbt)^2} f,
\quad
m_q = \frac{\lm_1^q \lm_2^q}{\sqrt{2}} \frac{f}{M_Q} v, \quad \hbox{ for }
q=t,s,d,\quad
Q=T,S,D,
\eeq
Small masses of $m_d$ and $m_s$ are
satisfied simply by the condition $\lm_1^{s,d} \ll \lm_2^{s,d}$,
which implies
\beq
\label{eq:small:mixing:D:S}
\theta_{D,S} \simeq 0.
\eeq
Accepting this simplification,
this model has the following five parameters:
\beq
\label{eq:model:parameters}
f, ~ m_\eta, ~  \tbt,  ~ x_{\lm} \left( \equiv \lm_1^t / \lm_2^t\right),  ~ M_D, ~  M_S.
\eeq
In Ref.\;\cite{cheung-song}, it is shown that proper EWSB prefers rather large $\tbt$
around 10.

%

Focused on $\eta$,
we put its Yukawa couplings as
\beq
\label{eq:Yukawa:eta:definition}
\Lg_Y^\eta =
- i \sum_f \frac{m_f}{v} y^\eta_f \eta \bar{f} \gm_5  f
+ \frac{m_t}{v}
\left(
i\eta \overline{T} P_R t + H.c.
\right)\,,
\eeq
where the index $f$ includes all of the SM fermions and heavy fermions,
$m_f$ is the fermion mass, $v$ is the Higgs VEV,
and $T$ is the heavy top partner.
We ignore $\eta$-$D$-$d$ and $\eta$-$S$-$s$ couplings due to
their small mixing angles in Eq.(\ref{eq:small:mixing:D:S}).
The parameter $y^\eta_f$ indicates the
ratio of the $\eta$ Yukawa coupling to the SM Higgs Yukawa coupling, given by
\bea
\begin{array}{rcl}
y^\eta_{l} &=& y^\eta_{d,s} = y^\eta_b  =-y^\eta_{u,c}
= -y^\eta_t= \dfrac{\sqrt{2} \, v}{ f} \cot 2 \beta \,,
\\
y^\eta_Q &=& - \dfrac{v}{f}
\left[ \cos 2 \beta  + \cos 2 \theta_Q  \right] \csc  2 \beta \,, \hbox{ for } Q=D,S,T\\
\end{array}
\eea
where $l=e,\mu,\tau$.

\section{$\eta$ Decay}
For $m_\eta < m_Z+ m_H$, the $\eta$ decays dominantly into a pair of
SM fermions that is kinematically allowed.
The decay rate of $\eta \to f\bar{f}$ is
\beq
\label{eq:Gm:eta:ff}
\Gm(\eta\to f\fbar) =
\frac{N_C}{8\pi }\left(\frac{m_f y^\eta_f}{v} \right)^2 m_\eta \lm^{1/2}_\eta
,
\eeq
where $\lm_\eta = 1-4 m_f^2/m_\eta^2$, and
$N_C$ is the color factor of the fermion $f$.
Since $\Gm(H \to f \bar{f})$ has
a factor of $\lm_H^{3/2}$, a pseudoscalar boson $\eta$ with mass
just above twice of a fermion mass has larger decay rate.
Since the decay rate is proportional to the fermion mass,
$\eta$ boson with $m_\eta < 2 m_t$ mainly decays into a $b\bar{b}$ pair.
In the following calculation of decay widths,
we use the running mass of the quarks evaluated at the scale $m_\eta$
to calculate the Yukawa coupling, but not in the phase space
factor.
This is why the partial width into
$c\bar c$ is  smaller than that into $\tau^+ \tau^-$.


As in the Higgs boson case,
the radiative decay rates of $\eta$ into $gg$ and $\gm\gm$
are also important.
Since $\eta$ has no coupling with the \emph{charged} gauge bosons,
the partial decay widths into $\gm\gm$ and $g g$ are, respectively,
\bea
\Gm(\eta \to g g) &=&
\frac{\alpha_s^2 m_\phi^3}{32 \pi^3 v^2}
\left|
\sum_f \frac{1}{2} y^\eta_f F^\eta_{1/2}(\tau_f)
\right|^2,
\\
\Gm(\eta \to \gm\gm) &=&
\frac{\alpha^2 m_\phi^3}{256\pi^3 v^2}
\left|
\sum_f
y^\eta_f N_{C}^f Q_f^2 F^\eta_{1/2}(\tau_f)
\right|^2,
\eea
where $\tau_f= 4 m_f^2/m_\eta^2$,
$N_C^f$ and $Q_f$ are, respectively, the color factor and the electric charge
of the fermion running in the loop.
The dimensionless loop factor $F^\eta_{1/2}(\tau)$ is
\beq
F^\eta_{1/2}(\tau) = -2 \tau f(\tau),
\eeq
with
\begin{equation}
    f(\tau) = \left\{ \begin{array}{lr}
        [\sin^{-1}(1/\sqrt{\tau})]^2, & \tau \geq 1, \\
        -\frac{1}{4} [\ln(\eta_+/\eta_-) - i \pi]^2, & \, \tau < 1,
        \end{array}  \right.
        \quad \eta_{\pm} = 1 \pm \sqrt{1-\tau}.
\end{equation}

\begin{figure}[t]
\centering
\includegraphics[width=3.5in]{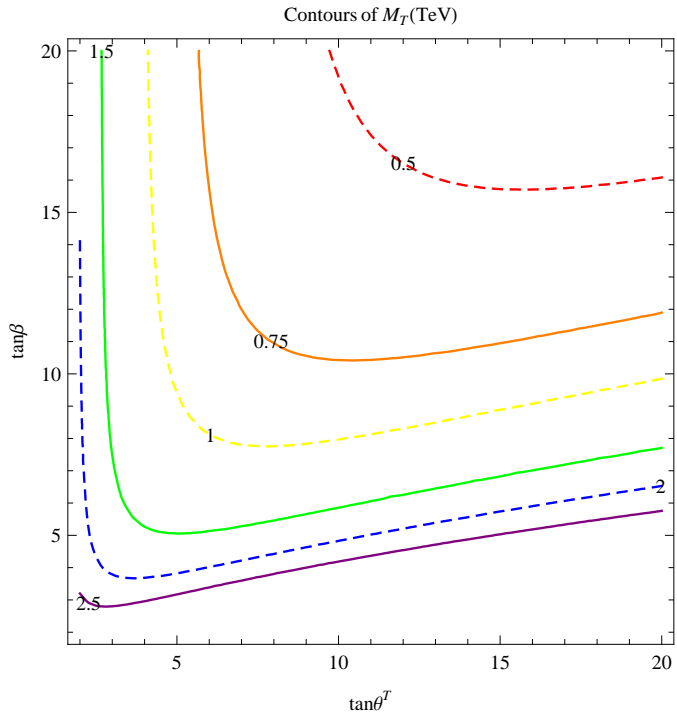}
\caption{\small
The dependence of $M_T$ (in unit of {\rm TeV}) on $\tan\beta \equiv \tbt$ and
$\tan\theta_{T}$. We have set $f=4$ {\rm TeV}.}
\label{fig1}
\end{figure}

\begin{figure}[t]
\centering
\includegraphics[width=3in]{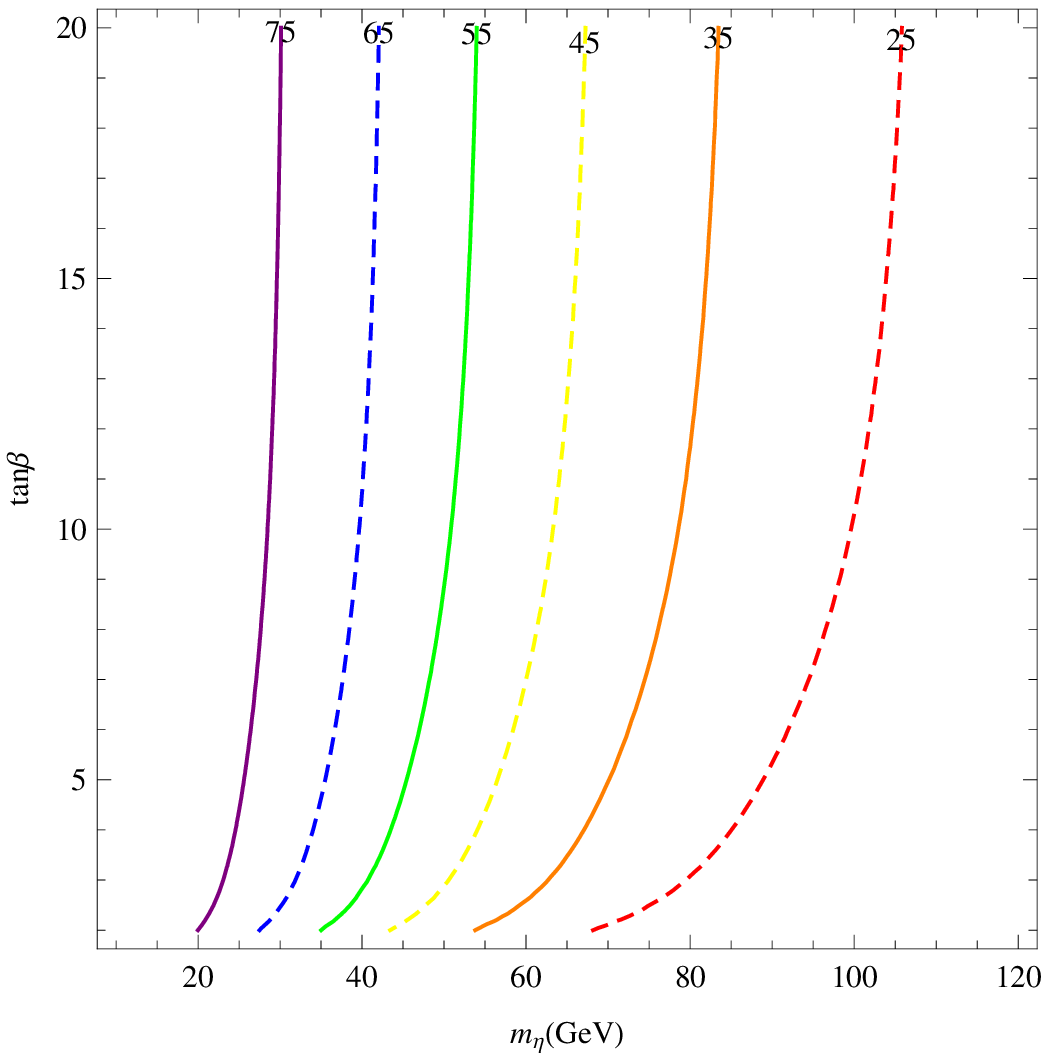}\phantom{xxx}
\includegraphics[width=3in]{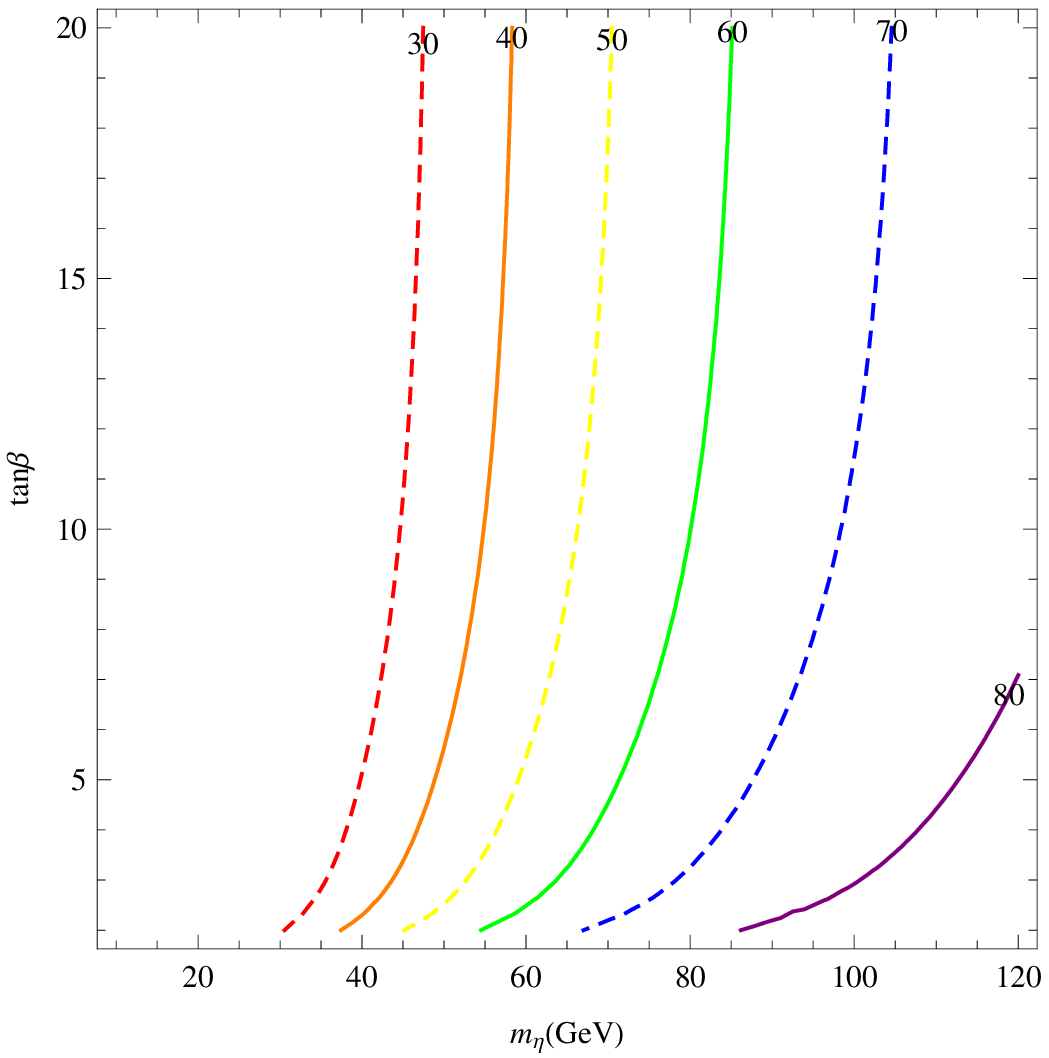}
\\ [-0.3 cm]
{(a) }   \hspace{3in}          { (b)}
\\[0.5cm]
\includegraphics[width=3in]{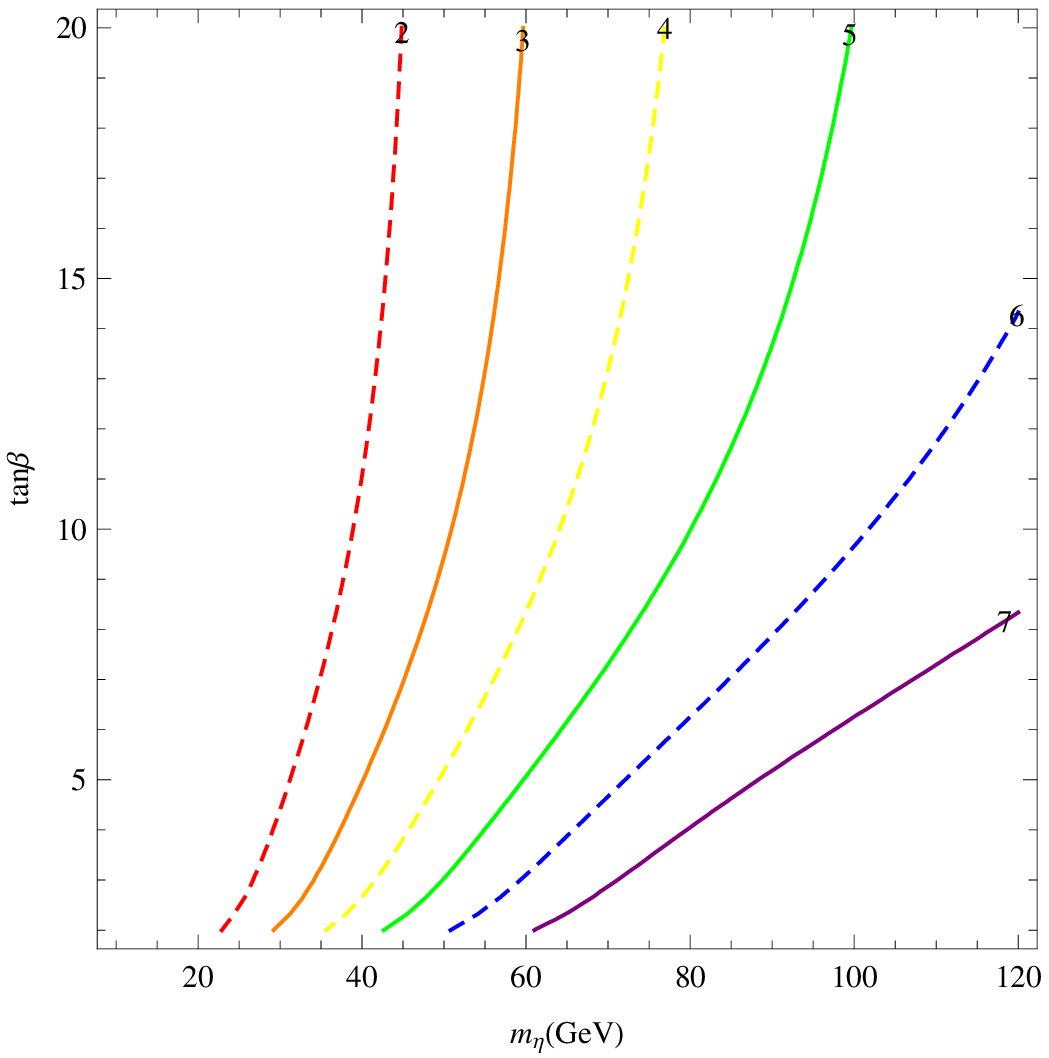}
\\ [-0.3 cm]
{(c)}
\caption{\small
Contour plots of Br$( \eta \to b \bar b) \times 10^2$ in (a),
Br$(\eta\to g g)\times 10^2$ in (b), and Br$(\eta\to \gamma \gamma)\times 10^4$ in (c)
on $m_\eta$   and $\tan\beta \equiv \tbt$
plane.  The parameter $f$ is fixed at $4$ TeV.
}
\label{fig2}
\end{figure}

For $m_\eta$ above the $b\bar b$ threshold the dominant
decay mode is into the $b\bar{b}$ pair.  However, due to huge QCD
background it is very difficult to identify the $\eta$ boson in this
mode, unless it is produced associated with some leptonic final
states.  The same difficulty is expected for $\eta \to gg$.
On the other hand, the decay
mode into $\tau^+ \tau^-$ could be useful, especially for $2\, m_\tau
< m_\eta < 2\, m_b$.  In this mass range, about 50\% branching ratio
is possible for $\eta \to \tau^+ \tau^-$. We shall concentrate on
this mode in Sec.\,V.

In Fig.\,\ref{fig1}, we show the dependence of $M_T$ on $\tbt$
and $\tan \theta_T$.
We have set $m_D = 1.53$ TeV and $m_S = 1.76$ TeV.
Since proper EWSB can be achieved by large $\tbt$ around 10,
the heavy top mass is somewhat sensitive to $\tan\theta_T$:
For $\tan\theta_T>5$, $M_T$ is relatively light below 1 TeV;
for $\tan\theta_T<5$, $M_T$ becomes heavier above 1 TeV.

In Fig.\,\ref{fig2} (a)--(c), we show the contours of
Br$(\eta \to b {\bar b})$, Br$(\eta \to g g)$,
and Br$(\eta \to \gamma \gamma)$, respectively,
in ($m_\eta$, $\tan \beta\equiv \tbt$) plane.
The presented value for $\eta \to b \bar b$ and $\eta \to g g$
is in unit of $10^{-2}$, while that for $\eta \to \gamma \gamma$
is in unit of $10^{-4}$.
We vary $m_\eta \in [20,~120]$ GeV,
and $\tbt \in [2,20]$.
The branching ratios are quite sensitive to $m_\eta$,
but relatively insensitive to $\tbt$.
As in Fig.\,\ref{fig1}, we used  $m_T>500$ GeV, $m_D = 1.53$ TeV and $m_S = 1.76$ TeV.

\begin{figure}[h]
\centering
\includegraphics[width=3.2in]{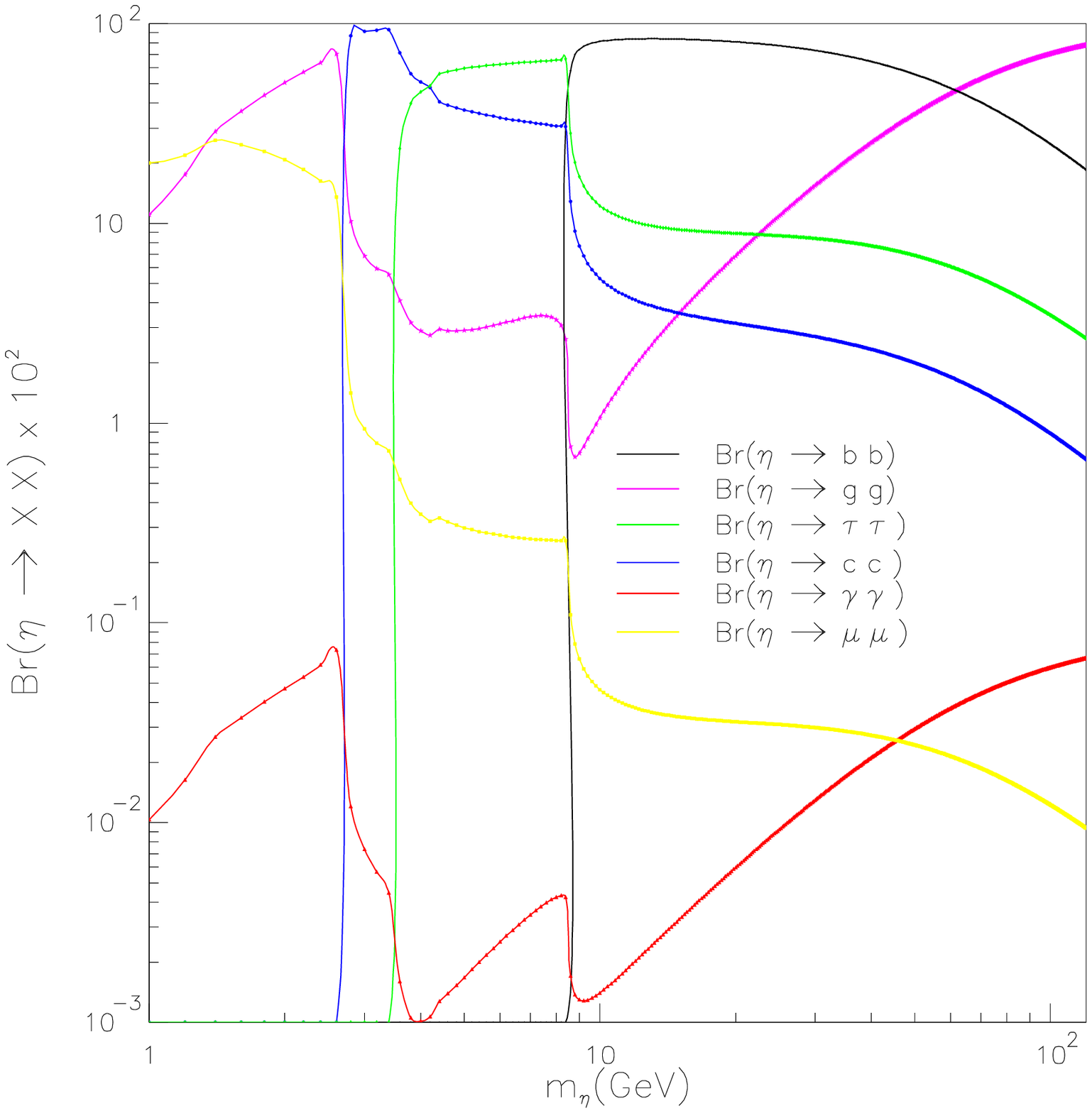}
\includegraphics[width=3.2in]{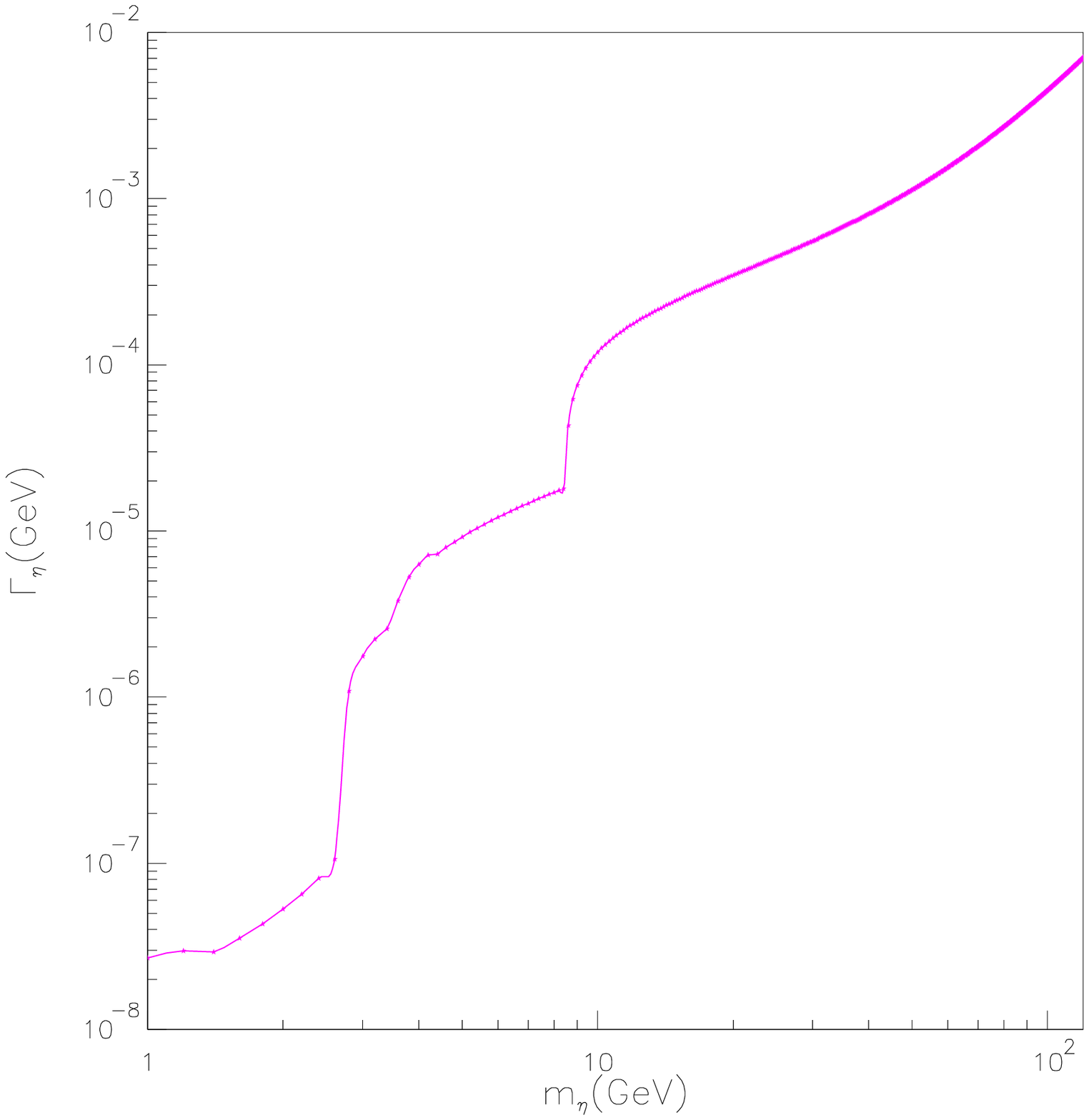}
\medskip
\hskip 0.4 cm { (a) } \hskip 6.3 cm { (b)}
\caption{\small
(a) Branching ratios and (b) total decay width of $\eta$ for
$1 \; {\rm GeV} < m_\eta < 120\; {\rm GeV}$. We fix
$\tbt =20$ and $f=4$ TeV.
}
\label{fig3}
\end{figure}

In Fig.\,\ref{fig3}, we show the branching ratios for the dominant
decay modes of $\eta$.
For $m_\eta < 2 \, m_c$  Br$(\eta \to g g) $ is the largest, whereas
in the range $2 m_c < m_\eta < 2 m_\tau$
Br$(\eta \to c \bar{c}) $ is the largest.
For $2 \,m_\tau < m_\eta < 2\, m_b$, the largest is Br$(\eta \to \tau^+ \tau^-) $, followed by
Br$(\eta \to c \bar{c}) $.
When $2\, m_b < m_\eta < 120$ GeV, Br$(\eta \to b \bar{b}) $ becomes dominant.
As $m_\eta$ further increases, however,
Br$(\eta \to g g) $ takes over.  This is due to
the enhancement from the contributions of heavy $T$, $D$ and $S$,
 which are non-decoupled in the triangle loops.
Such an enhancement is helpful for $\eta$ production at the
LHC via gluon fusion.
On the other hand, Br$(\eta \to \gamma \gamma)$
is only at the level $10^{-4}$ in most of the parameter space.
This is due to the absence of the $\eta$-coupling with charged gauge bosons.

\section{$\eta$ production at the LHC}

Main production channels for $\eta$ at the LHC are, in the order
of the size of cross sections,
\begin{enumerate}
    \item gluon fusion: $g g \to \eta$;
    \item $b\bar{b}$ fusion: $b \bar{b} \to \eta$;
    \item associated production with $H$:
 $q\bar{q} \to Z, Z', \im (Y^0) \to H \eta$;
    \item associated production with $t\bar{t}$:
        $gg, q\bar{q} \to t\bar{t} \eta$; and
    \item decay from $T$: $T \to t \eta$ and $\bar T \to \bar t \eta$.
\end{enumerate}
The resulting cross sections are to be
compared in subsection F.

\subsection{Gluon fusion}
For gluon fusion the cross section at $p p$ hadron collider
with the c.m. energy $s$ $(\sqrt{s} =14\tev$ for the LHC) is
\beq
\sigma (g g \to \eta) = \int_{\tau_\eta}^1 \frac{d x}{x}
f_{g/p}(x) f_{g/p}\left( \frac{\tau_\eta}{x} \right)\, \tau_\eta \,\sigma_0(gg\to\eta),
\eeq
where $\tau_\eta = m_\eta^2/s$,
$f_{g/p}$ is the parton distribution function of a gluon inside a proton,
and
\beq
\sigma_0(gg\to\eta) = \frac{\pi^2}{8 m_\eta^3} \Gm(\eta \to g g).
\eeq

\subsection{$b\bar b$ fusion}
For the $b\bar{b}$ fusion the cross section is
\bea
\sigma (b \bar{b} \to \eta)
&=& \int_{\tau_\eta}^1
 \frac{d x}{x}
\left[
f_{b/p}(x) f_{\bar{b}/p}\left( \frac{\tau_\eta}{x} \right)
+
f_{\bar{b}/p}(x) f_{b/p}\left( \frac{\tau_\eta}{x} \right)
\right]\, \tau_\eta \,\sigma_0(b \bar{b} \to \eta)
\\ \no
&=&
2 \int_{\tau_\eta}^1
 \frac{d x}{x}
f_{b/p}(x) f_{\bar{b}/p}\left( \frac{\tau_\eta}{x} \right)
\, \tau_\eta \,\sigma_0(b \bar{b} \to \eta),
\eea
where
\beq
\sigma_0(b \bar{b} \to \eta) = \frac{4 \pi^2}{ 9 m_\eta^3} \,
\Gm(\eta \to  b\bar{b}).
\eeq


\subsection{$H\eta$ associated production}
The process of $q \bar{q}  \to \eta H$ is mediated by $Z$, $Z'$,
and $\im(Y^0)$ gauge bosons.
Since the gauge coupling of $Y^0$ with the SM fermion is
suppressed by $v/f$ and $1/\tbt$,
we ignore it in the following.
The interaction Lagrangian is parameterized by
\beq
\Lg = - g_Z \sum_{i,q} \bar{q} \gm_\mu
\left[
{g}^{q}_{iR} P_R
+
{g}^{ q}_{iL} P_L
\right] Z_{i}^{\mu}
+ \sum_{i,q} c_i \left[ H  \partial_\mu \eta  -\eta \partial_\mu H \right]Z_i^\mu
\,,
\eeq
where $g_Z = g/c_W$, $i=1,2$, $Z_{1,2}=Z, Z'$, and
\bea
\label{eq:gRL}
{g}^{q}_{1R} &=& -x_W Q_q, \quad {g}^q_{1L} = \left( T^q_3 \right)_L - x_W Q_q,
\\ \no
{g}^{q}_{2R} &=& \frac{Q_q x_W}{\sqrt{3-4 x_W}} ,
\quad
{g}^q_{2L} = \frac{1}{\sqrt{3-4 x_W}}
\left(
-\frac{1}{2} +\frac{2}{3}x_W
\right),
\quad
q=u,d,c,s,
\\ \no
{g}^{b}_{2R} &=& -\frac{1}{3}\frac{ x_W}{\sqrt{3-4 x_W}} ,
\quad
{g}^b_{2L} = \frac{1}{\sqrt{3-4 x_W}}
\left(
\frac{1}{2} -\frac{1}{3}x_W
\right),
\\ \label{eq:ci}
c_1 &=& \sqrt{2} \left( \tbt - \frac{1}{\tbt} \right)
\frac{m_Z}{f},
\quad
c_2 =- \sqrt{2}
\left( \tbt - \frac{1}{\tbt} \right)
\frac{m_Z}{f} f_{Z'}.
\eea
Here $x_W =\sin^2 \theta_W$ and $f_{Z'}= \cw (1-\tw^2)/\sqrt{3-\tw^2} \approx 0.682 $.

For the process of
\beq
q (p_1) + \bar{q}(p_2) \to \eta(k_1) + H(k_2),
\eeq
the parton level differential cross section is
\beq
\frac{d \hat{\sigma}}{d \cos\theta^*}(q \bar q' \to \eta H) =
\frac{\lambda^{1/2}\left(m_\eta^2/\shat,m_H^2/\shat \right)}{36 \pi \shat}
\frac{1}{3}(|\hat{g}_{qL}|^2 + |\hat{g}_{qR}|^2) (1-\cos^2\theta^*),
\eeq
where $\theta^*$ is the scattering angle of $\eta$ with respect to the incoming quark $q$
in the parton c.m. frame,
$\shat=(p_1+p_2)^2$,
and $\lambda(a,b)=1+a^2+b^2-2a-2b-2ab$.
The effective couplings $\hat{g}_{qL,qR}$ are
\beq
\hat{g}_{qX} = \sum_{i=1,2}\; \frac{ \shat }{\shat -m_i^2+ i m_i \Gm_i}\,
 g_Z\,
c_i \, g^q_{iX},
\qquad \hbox{for } X=R,L.
\eeq
The cross section of $pp$ collision is then
\bea
\sigma(pp\to \eta H) &=&
\int d x_A d x_B
\biggl[
\left\{
f_{u/p}(x_A)f_{\bar{u}/p}(x_B) + f_{c/p}(x_A)f_{\bar{c}/p}(x_B)
\right\} \hat{\sigma}( u \bar{u} \to \eta H)
 \\ \no
 &&
+
\left\{
f_{d/p}(x_A)f_{\bar{d}/p}(x_B) + f_{s/p}(x_A)f_{\bar{s}/p}(x_B)+ f_{b/p}(x_A)f_{\bar{b}/p}(x_B)
\right\} \hat{\sigma}( d \bar{d} \to \eta H)
\\ \no &&
+ (A \leftrightarrow B)
\biggl]
\eea
where $f_{a/A}$ is the parton density of $a$ inside the hadron $A$.

\subsection{$t\bar t \eta$ associated production}
There are two contributing subprocesses:
\[
  q \bar q \to t \bar t \eta, \qquad g g \to t \bar t \eta \;.
\]
The latter dominates at the LHC energy because of large gluon luminosity.
We write down the helicity amplitudes in the appendix, and use FORM to
evaluate the square of the amplitudes.

\subsection{In the $T$ decay $T \to t \eta$}

A pair of $T \bar{T}$ is produced by QCD interactions, similar to a
top-quark pair.  When the $T$ is heavy enough, single-$T$ production
is kinematically advantageous \cite{SingleT}.
In little Higgs models, the heavy $T$ quark decays dominantly into $tH$,
$tZ$, and $bW$\;\cite{smoking}.
When neglecting final-state masses over $M_T$,
the partial decay rates are
\beq
\Gm( T \to t H) = \Gm(T \to t Z) = \frac{1}{2}\,\Gm(T \to b W) = \frac{\lm_T^2}{32\pi} M_T,
\eeq
where
\beq
\lm_T = \frac{\tbt}{\tbt^2 + 1} \left(\xlm - \frac{1}{\xlm} \right) \frac{m_t}{v}
\,.
\eeq
Here $\xlm =\lm_1^t/\lm_2^t$.
If these are the only decay modes of $T$,
the branching ratios of $T$ would show simple relation of
$\br( T \to t H):\br(T \to t Z) : \br(T \to b W)=1:1:2$.
However the SLH model allows another important decay mode
of $T \to t \eta$.
Its partial decay rate is
\beq
\Gm( T \to t \eta ) = \frac{1}{32\pi} \frac{m_t^2}{v^2} M_T.
\eeq
Since EWSB prefers large $\tbt$ so that $\lm_T$ is suppressed if $\xlm$ is not as large as $\tbt$,
this decay mode can be important.
For example,
two benchmark points in Ref.\cite{cheung-song-yan}
have sizable decay rate of $T \to \eta t$:
$\br(T \to \eta t)\approx 45\%$ in the SHL$\mu$-A case
and $\br(T \to \eta t)\approx 21\%$ in the SHL$\mu$-B case.

\begin{figure}[t!]
\centering
\includegraphics[width=5in]{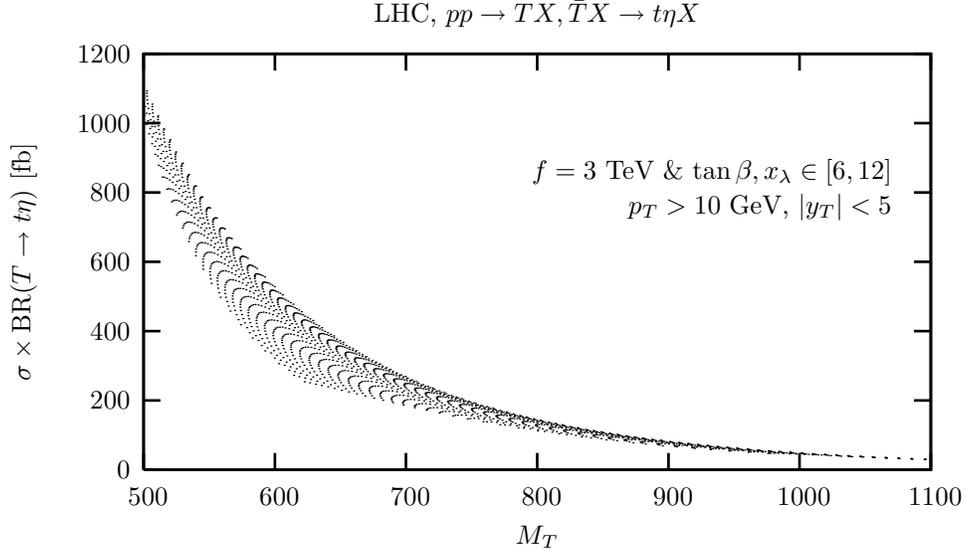}
\caption{\small
Sum of cross sections of $pp \to TX$ and $pp\to \bar{T}X'$,
multiplied by the branching ratio of $T\to t\eta$ in units of fb.
We fix $f=3\tev$, while varying $\tbt \in [6,12]$
and $\xlm \in [6,12]$.
}
\label{fig:Tdecay:sigbr:dot}
\end{figure}

In Fig.\,\ref{fig:Tdecay:sigbr:dot}, we show the production cross
section of a single heavy top $T$ at the LHC, multiplied by its branching
ratio for $T \to t \eta$.  We fix $f=3\tev$  and vary
$\tbt$ and $\xlm$ in the range between 6 and 12.
We have included the single-$T$ production as well as
the single-$\bar{T}$ production.  As stressed in Ref.\cite{SingleT},
the non-negligible $2 \to 3$ process of $q g \to T \bar{b}q'$ with its
charge-conjugated production is also included.
Only for very optimal case of $M_T$
around 500 GeV, the $\eta$ production from the $T$ decay can reach 1
pb.

\subsection{Comparisons}

\begin{figure}[t]
\centering
\includegraphics[width=4in]{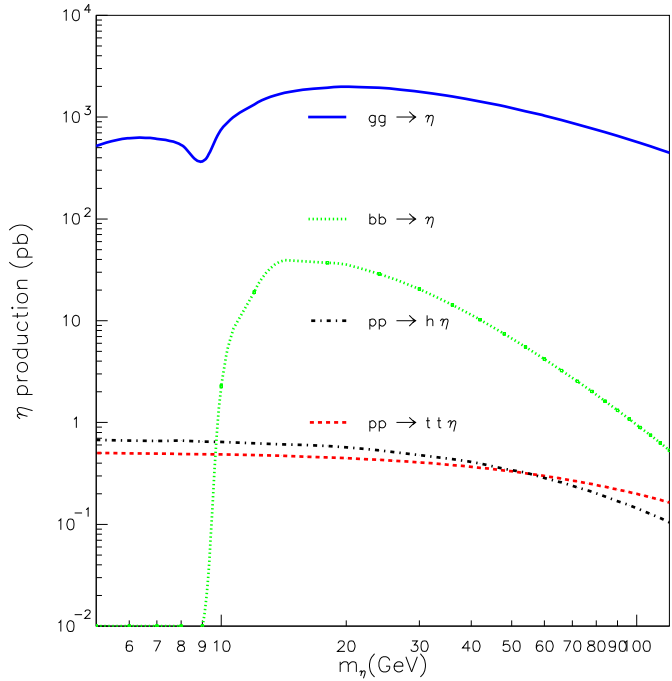}
\caption{\small
Various production cross sections for $\eta$ at the LHC
in the mass range  $5 \gev < m_\eta < 120 \gev$. We fix
$\tan \beta \equiv \tbt =20$ and $f=4 \tev$.
}
\label{fig5}
\end{figure}

In Fig.\,\ref{fig5}, we show the production cross sections against
the mass of the $\eta$ for various production channels at the LHC.
The most dominant production channel is the gluon fusion,
followed by the $b\bar b$ fusion.
In both channels, there is only $\eta$ in the final state, which will
decay into either a $b\bar b$ pair or $\tau^+ \tau^-$ pair depending on
$m_\tau$.  Both modes suffer, respectively,
from large QCD background and the Drell-Yan background.
We do not consider these two
production channels in the subsequent sections.
In the associated production of $h \eta$, one has the addition $h\to b\bar b$
decay to put the handle on.  We will study this channel in
detail in the next section.  Another channel of interest is the
$t\bar t \eta$.  We will consider the signal-background in Sec.\,VI.
Finally, one should not ignore the $\eta$'s from the decay of the heavy $T$.
As long as $T$ is not too heavy such that its production rate is sizable, we
expect enough $\eta$'s from $T$ decays.

\section{The signal and background of
the process $ p p \to h \eta \to b {\bar b} \tau^-  \tau^+ $}

\begin{figure}[t]
\centering
\includegraphics[width=3.2in]{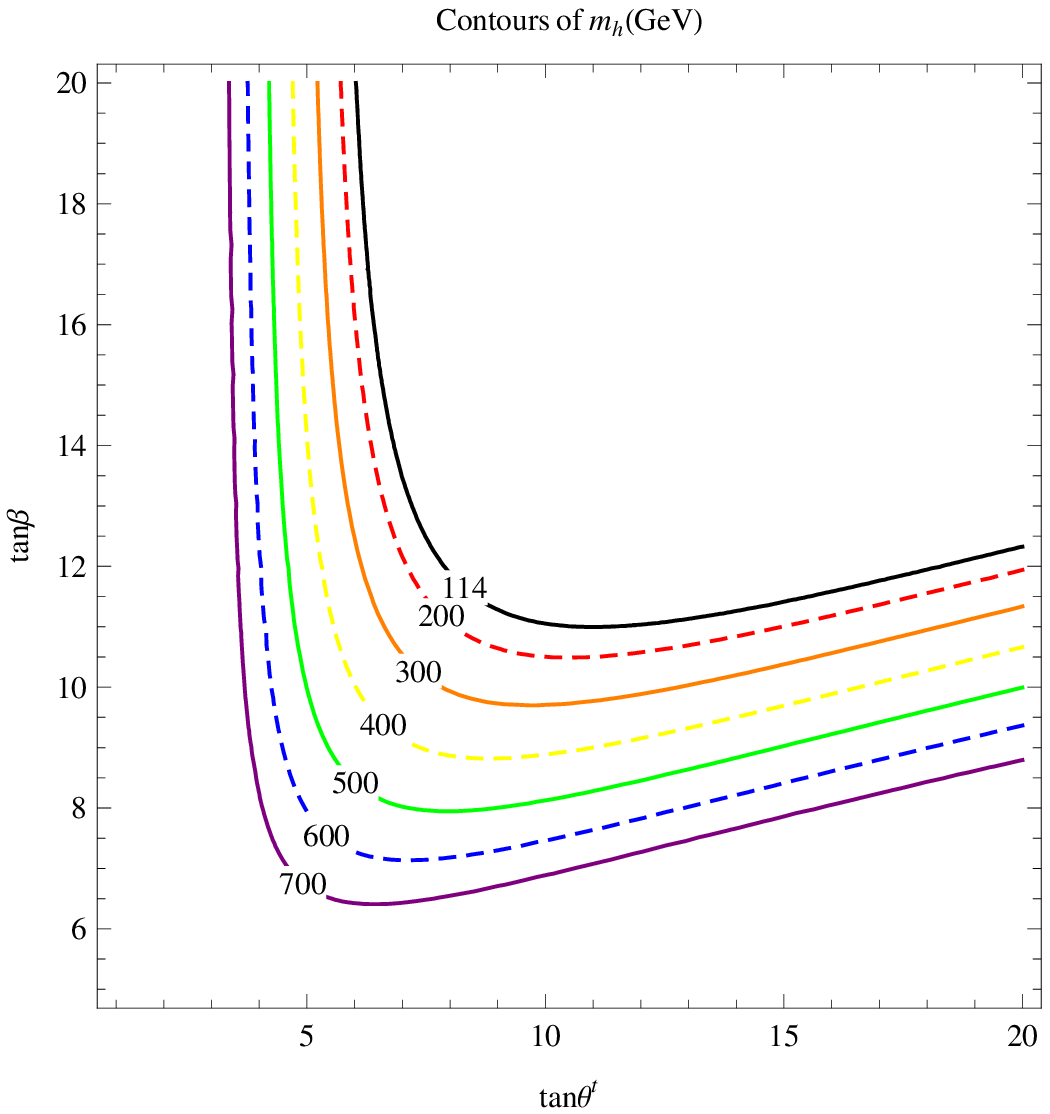}
\includegraphics[width=3.2in]{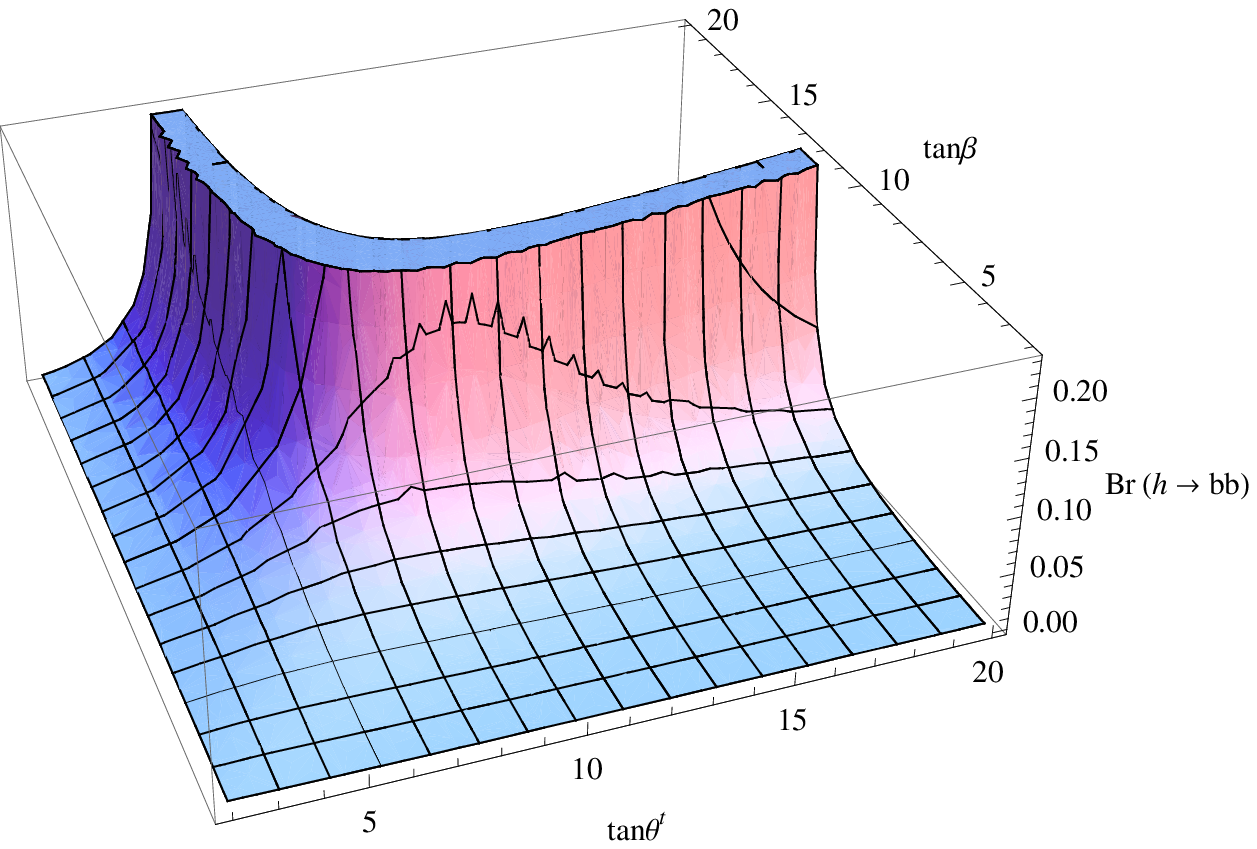}
\vskip -0.05cm \hskip 0.4 cm {(a) } \hskip 5.8cm {(b)}
\caption{\small
(a) The mass of the Higgs boson in units of GeV and (b)
Br$( h \to b {\bar b})$ in the two-dimensional
$\tan\theta_T-\tbt$ plane. We fix $f=4\tev$.
}
\label{fig6}
\end{figure}

\begin{figure}[t]
\centering
\includegraphics[width=3.2in]{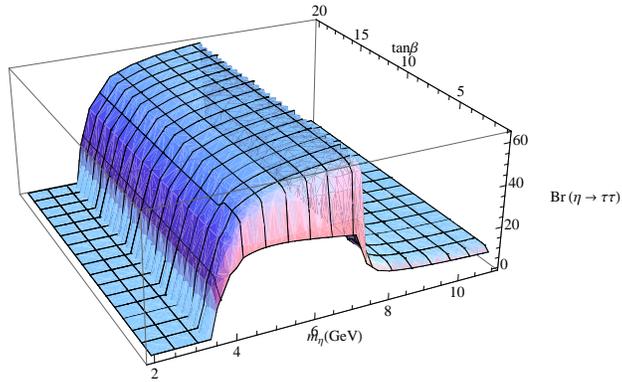}
\caption{\small
The branching ratio Br$(\eta \to \tau^+ \tau^-)$ in unit of $10^{-2}$
for $f=4\tev$.
}
\label{fig7}
\end{figure}

In this section, we study the signal and backgrounds of the
$\eta$ production associated with the Higgs boson.
To enhance the signal, we consider the following mass ranges of $h$ and $\eta$
where the subsequent decay of $h \to b\bar{b}$
and $\eta \to \tau^+ \tau^-$ are important:
\beq
\label{eq:mass:range}
2 m_\tau < m_\eta < 2 m_b,
\quad 114 \; \textrm{GeV} < m_h < 2 m_W\,.
\eeq
Then $ p p \to h \eta$ production has
the dominant
decay mode of $b\bar b \tau^+ \tau^-$ without missing transverse energy.
The kinematic cuts imposed on $b$'s and $\tau$'s  are
\bea
p_T(j) > 15\; \gev, \quad
p_T(\tau) > 10\; \gev, \quad |y_{j,l}|<2.5\,, \quad \Delta R > 0.4\,.
\eea
The reconstruction efficiencies of $b$ and $\tau$ are taken as $0.5$
and $0.4$, respectively.  The rejection rates of gluon and light quarks
faking $\tau$ are taken as $300$, while the rejection rate of $c$ quark
faking $\tau$ is taken as $20$\;\cite{tau}.
To simulate the detector effects, we smear the energies of the jets and
$\tau$'s with the following Gaussian resolutions
\bea
\frac{ \Delta E_j }{E_j} &=& \frac{50\%}{\sqrt{E_j}} \oplus 3 \%\,, \\
\frac{ \Delta E_\tau }{E_\tau} &=& \frac{10\%}{\sqrt{E_\tau}} \oplus
0.7 \%\,,
\eea
where $E_{j,\tau}$ are measured in GeV.
We also introduce the following invariant mass cuts on $m_{b\bar b}$
and $m_{\tau\tau}$ to suppress the backgrounds:
\bea
m_h - 10 \;{\rm GeV} < m_{b\bar b} < m_h  + 10 \; {\rm GeV} \,, \nonumber \\
m_\eta - 2 \; {\rm GeV} < m_{\tau\tau} < m_\eta + 2 \; {\rm GeV} \,.
\label{mcuts}
\eea
To further suppress the background from
$t {\bar t} \to b\bar b \tau^+\tau^- + {E\hspace{-2.5mm}\slash }$, we reject
events with the missing transverse energy
${E\hspace{-2.5mm}\slash }_T > 50$ GeV, which
is about the threshold of the missing transverse energy
signature of CMS and ATLAS detectors. We observe that this cut can reduce the
$t {\bar t}$ background by $2/3$.
For Fig.\,\ref{fig6}, we present the Higgs mass and its branching ratio into $b\bar b$
in the plane of $\tan\theta_T$ and $\tbt$
with fixed $f=4\tev$. The branching ratio
Br$(\eta \to \tau^- \tau^+)$ is given in Fig.\,\ref{fig7},
which is almost insensitive to the value of $\tbt$.

In order to simulate the signal,
we select the following benchmark point:
\beq
\label{eq:benchmark}
f=4  \tev, \quad
\tan\theta_T=6.03, \quad \tbt=20, \quad
m_\eta=8 \gev.
\eeq
The relevant physics quantities corresponding to this benchmark point are
$\Gamma_\eta=3.8\times 10^{-5}$ GeV, Br$(\eta \to \tau^+
\tau^- )=38.5\%$, $m_h=115$ GeV, $\Gamma_h=6.6 \times 10^{-3}$ GeV,
Br$(h \to b {\bar b})=88.7\%$ and Br$(h \to \eta \eta)= 0.43\%$. The
$Z$-$h$-$\eta$ coupling in Eq.\,(\ref{eq:Z:h:eta}) is about $0.312$ at the benchmark point.
The small branching ratio of $h \to \eta \eta$ is due to the small
$m_\eta$. At this benchmark point, the $Z^\prime$ is quite
heavy and does not appreciably contribute to $pp \to h \eta$.

We have several comments on the decay properties of $h$ and $\eta$ in
the mass region $2 m_\tau < m_\eta < 2 m_b$.
\begin{enumerate}
\item
  The main decay
  channels of $\eta$ are  $\eta \to \tau^+ \tau^-$ and
  $\eta \to c {\bar c}$.
  The branching ratio of $\eta \to \tau^+ \tau^-$ is the
  largest due to the fact that QCD corrections to Br$(\eta\to c\bar c)$
  encoded in the running mass $m_c$ make it substantially
  smaller than that of Br$(\eta \to \tau^+ \tau^-)$,
  even though the pole mass of charm quark is close to tau pole mass and
  this channel has a color factor of 3.
  In this mass region, the branching ratio
  Br$(\eta \to \tau^+ \tau^- )$ is not sensitive to the parameter
  $\tbt$.
\item The mass of
  Higgs is not sensitive to the change of $m_\eta$ due to the
  smallness of $m_\eta$.
\item The decay mode of $h\to \eta\eta$ is small unlike the case in
  Refs.\;\cite{cheung-song,cheung-song-yan}, because here the mass of
  $\eta$ is small and the decay rate is proportional to $m_\eta^4$.
\end{enumerate}

\begin{figure}[t]
\centering
\includegraphics[width=3.2in]{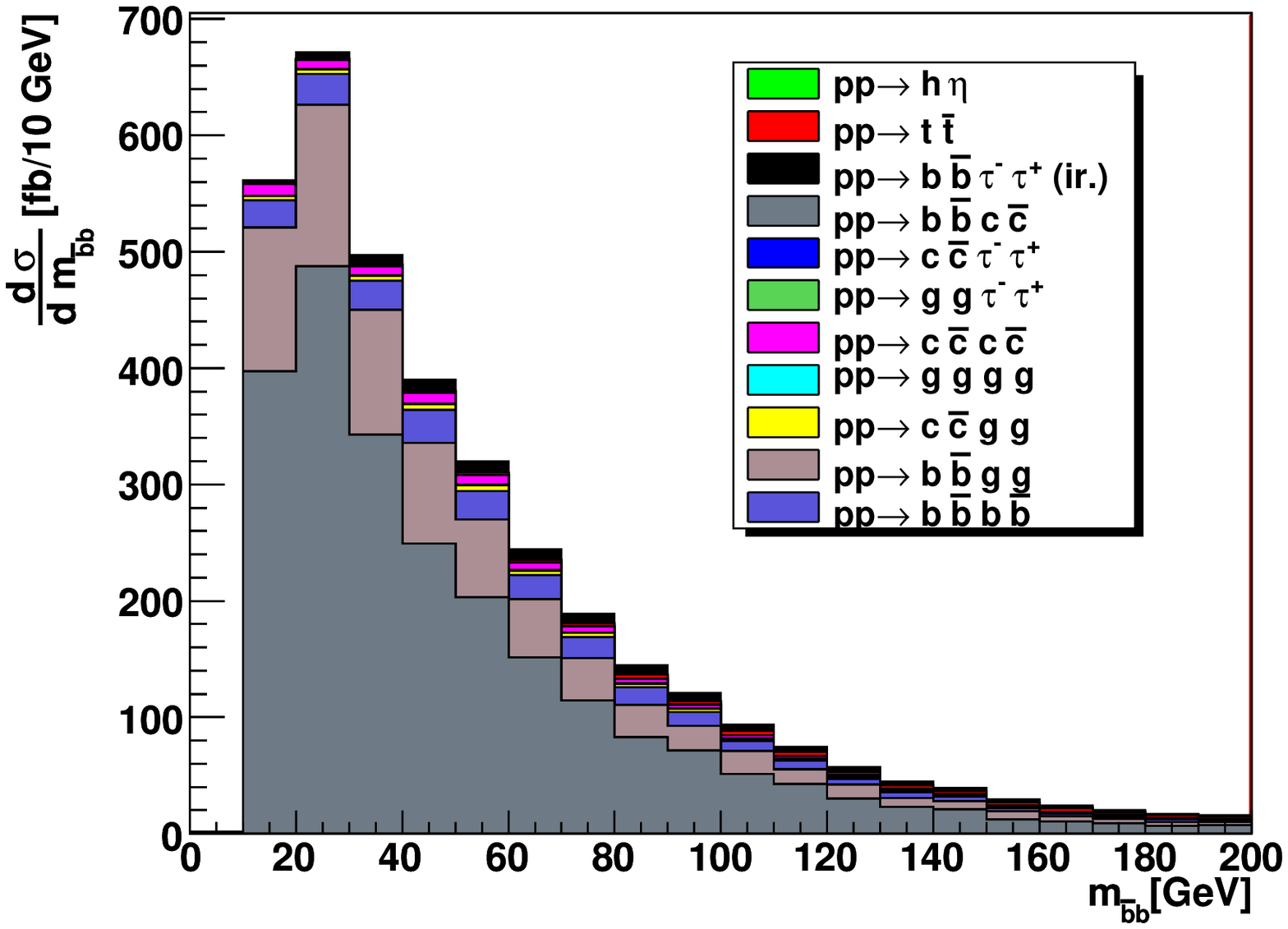}
\includegraphics[width=3.2in]{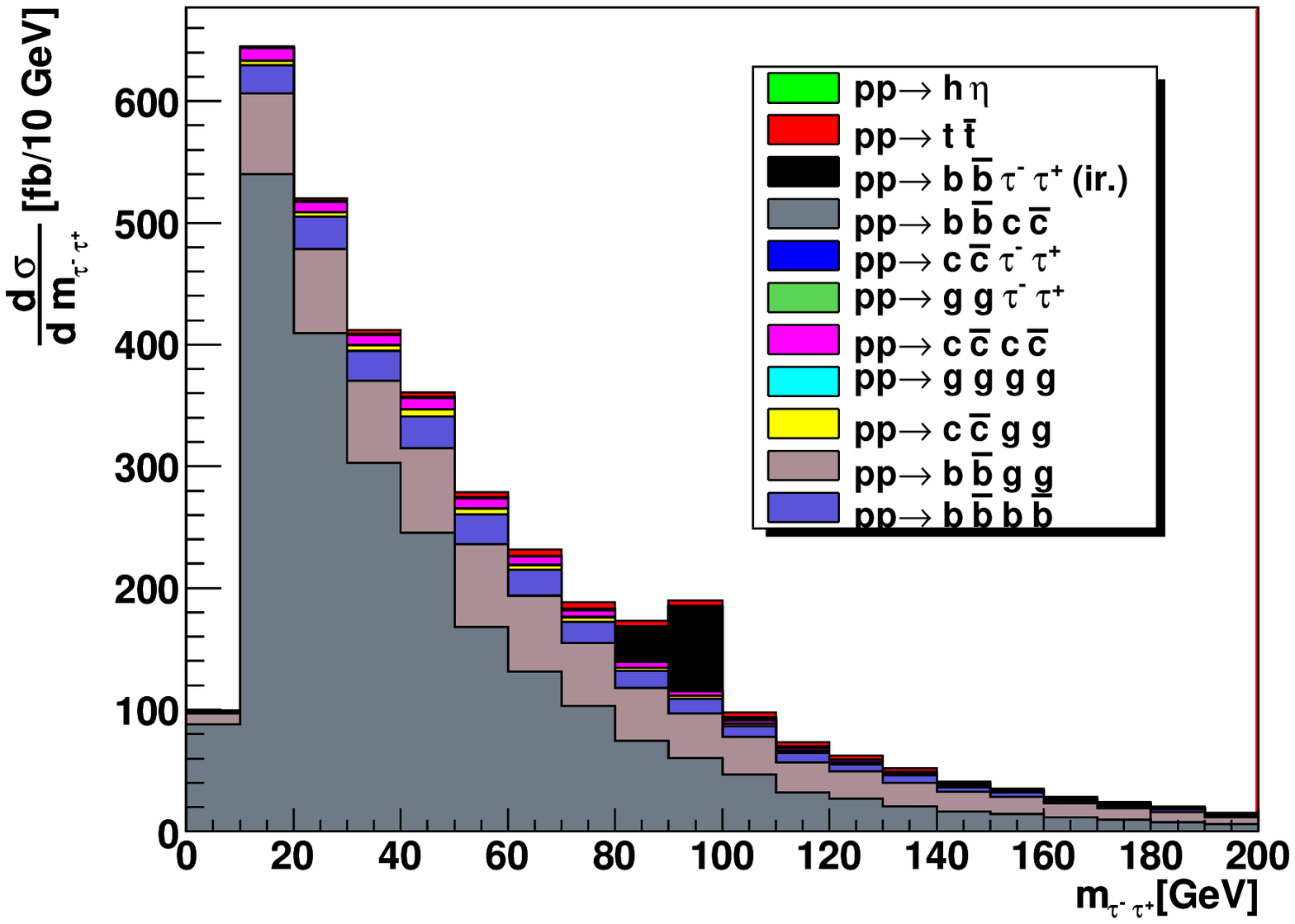}
\vskip -0.05cm
\hskip 0.4 cm { (a) } \hskip 5.8 cm {(b)}
\caption{\small
The invariant mass distributions of (a) $m_{b\bar b}$ and (b)
$m_{\tau\tau}$ for the signal and various backgrounds in
$pp \to h \eta \to b {\bar b} \tau^- \tau^+$ after imposing the
selection cuts.}
\label{fig8}
\end{figure}

In Fig.\,\ref{fig8}, we show the signal and various backgrounds
in both $m_{b\bar b}$ and
$m_{\tau \tau}$ invariant mass distributions. It is clear that the signal
is buried under the background and can hardly be seen.  From
these two plots, we
can identify that the $pp\to b {\bar b} c {\bar c} $ is the most dominant
background.
In order to separate the signal of $pp \to h \eta$ it is crucial to
enhance the rejection factor for $c$ quark faking $\tau$.
The final signal and background rates
are tabulated in Table \ref{bkgd-sgnt}.
\begin{table}
\begin{ruledtabular}
\begin{tabular}{| c | c | }
 Background processes & Cross Section (fb)  \\ \hline
$b{\bar b} c{\bar c}$  &   $ 1.87 $       \\
$b{\bar b} g g $  &   $ 0.38$       \\
$c{\bar c} g g$  &   $0.06$          \\
$c{\bar c} c{\bar c}$  &   $0.05$          \\
\hline
\hline
 Signal processes & Cross Section (fb)  \\ \hline
$b {\bar b} \tau^- \tau^+ $  &   $0.33$          \\
\end{tabular}
\end{ruledtabular}
\caption{ Background and signal cross sections in $pp \to h \eta
\to b\bar b \tau^+ \tau^-$ after imposing all the cuts and tagging
efficiencies.}
\label{bkgd-sgnt}
\end{table}

A few more comments on the signal are in order here.
\begin{enumerate}
\item The raw cross section of $p p \to h \eta$ is around $1$ pb.
  With the branching ratios of Br$(h \to b {\bar b})$ and Br$(\eta \to
  \tau^- \tau^+)$, the signal is around $300$ fb. After
  imposing the cuts on $\Delta R(b, {\bar b})$, $\Delta R(\tau^+,
  \tau^-)$, $y_{b}$ and $y_{\tau}$, the signal is reduced to $90$
  fb. In addition, the $p_T$ cuts on $b$'s and $\tau$'s further reduce the
  signal to $9.6$ fb, because most of events are lost due to the $p_T$
  cut on the not-so-energetic $\tau$'s  and the $\Delta R(\tau^-, \tau^+)$
  cut in each signal event. Considering the reconstruction
  efficiencies and the invariant mass cuts
  given in Eq.\,(\ref{mcuts}), the final reconstructed cross section is
  $0.33$ fb.
\item When we fix $m_\eta$ and increase $m_h$, we find that the final
  reconstructed cross section decreases. For instance when we take
  $m_h = 155$ GeV, the reconstructed cross section is only $0.04$
  fb. This reduction is not only due to the decrease of the cross
  section $\sigma(p p \to h \eta)$ but also due to the onset of
  the mode $h\to W W^*$.
\item Unlike MSSM and NMSSM, there is no dramatic enhancement
due to large $\tbt$ for the $pp\to h\eta \to b\bar{b}\tau^+\tau^-$
than the chosen benchmark point in Eq.(\ref{eq:benchmark}).
First the
  parameter $\tbt$ has an upper bound by the validity of
  perturbation expansion\;\cite{cheung-song}. Second,
  the coupling of $Z$-$h$-$\eta$ is around
  half of $g_Z$, which is almost the maximum value in the allowed
  parameter space. Third, the
  branching ratio of Br$(\eta \to \tau^- \tau^+)$ cannot change
  drastically with the increase of $\tbt$.
  This is to be compared with the MSSM and NMSSM model cases
  where both the $Z$-$h$-$\eta$ coupling and the branching ratio
  of Br$(\eta \to \tau^- \tau^+)$ might be enhanced by factors of $1.5$ and
  $1.5$, respectively. 
\end{enumerate}

\section{The signal and backgrounds of
$ p p \to t\bar t \eta$}
In this section we study the signal and backgrounds of
$ p p \to t\bar t \eta$, focused on
the benchmark point in Eq.\,(\ref{eq:benchmark}).
For simplicity  we first assume that the reconstruction efficiency of the
top quark is $100$ percent and can be fully reconstructed.
We can focus on the analysis of the signal and the most direct backgrounds
at the $t$ and $\bar t$ level.
The majority of backgrounds comes from $t\bar t jj$ with $jj = c \bar c,\;
gg,\;q \bar q$, among which $t\bar t c \bar c$ is the most serious background.
This is because of poor rejection factor of the charm quark, which
behaves similarly to a $\tau$-jet.  Another source of background
is $t\bar t \tau^+ \tau^-$, which is irreducible but essentially
small because the $\tau$'s are produced mostly off a virtual photon or
a $Z$ boson.

\begin{figure}[t]
\centering
\includegraphics[width=3.2in]{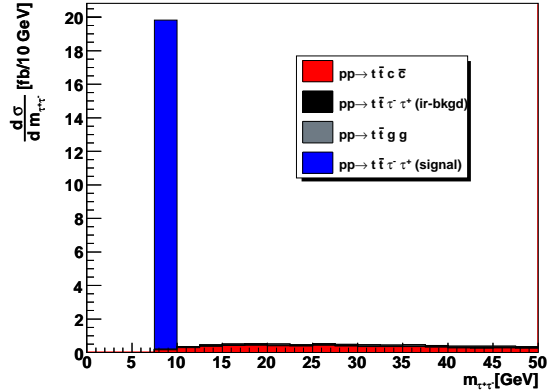}
\caption{\small
The invariant mass distributions of $m_{\tau\tau}$ for the signal
and various backgrounds in
$pp \to t\bar t \eta \to t {\bar t} \tau^- \tau^+$ after imposing the
selection cuts.}
\label{fig9}
\end{figure}

We note that the invariant mass cut on $m_{\tau\tau}$ is very crucial
for signal event selection.  It removes most of irreducible backgrounds,
in which the two $\tau$'s are emitted from a virtual photon, $Z$,
or a Higgs boson. It also suppresses the $t\bar tc {\bar c}$ background
effectively.
In Fig.\,\ref{fig9}, we show the spectrum of $m_{\tau \tau}$ for both
the signal and the major backgrounds.  It is obvious that the signal
clearly stands out of the backgrounds.

\begin{table}
\begin{ruledtabular}
\begin{tabular}{| c | c | }
 Background processes & Cross Section (fb)  \\ \hline
$t{\bar t} c {\bar c}$  &   $ 0.17$       \\
$t{\bar t} \tau^- \tau^+ $  &   $ 0.02$       \\
$t{\bar t} g g$  &   $0.02$          \\
\hline
\hline
 Signal processes & Cross Section (fb)  \\ \hline
$t{\bar t} \tau^- \tau^+$  &   $19.6$          \\
\end{tabular}
\end{ruledtabular}
\caption{ Cross sections for the signal and
various backgrounds in $pp\to t {\bar t} \tau^- \tau^+$
after imposing all cuts
\label{tttata-bkgd-sgnt}
}
\end{table}

In Table \ref{tttata-bkgd-sgnt}, we show the cross sections of major
backgrounds and signal after cuts.
Here we comment on the signal in more detail.
The total cross section of the $pp\to t\bar t \eta$ process is about
$500$ fb. After imposing the
kinematic cuts on the $\tau$'s, the cross section is
reduced to about $120$ fb.  Including the
reconstruction efficiency of $\tau$'s, it further reduces to about $19$ fb.
Note that we have only analyzed down to the $t$ and $\bar t$ level.
If we take the top-quark decays into account, there should be
more backgrounds such as $W^+ W^- +\,$jets and $W^\pm +\,$jets.  However, these
backgrounds are electroweak in nature and thus much smaller than the QCD
production of $t\bar t c \bar c$.

The signal-to-background ratio seems quite promising.
However, we still need to apply the top quark reconstruction and identification
efficiencies.
For a simple estimate of the top quark reconstruction,
we should include the reconstruction efficiency of two $b$ jets and
two $W$'s.
With one of the $W$'s decaying leptonically ($e$ and $\mu$)
while the other one decaying hadronically,
the reduction factor is therefore
\bea
f = e_b^2 \times 2 \times \frac{6}{9} \times \frac{2}{9}  \,.
\eea
where $e_b \approx 50\%$ is the $B$-tagging efficiency.
Thus, the more realistic reconstructed cross section of
$p p \to t {\bar t} \tau^- \tau^+$ is estimated to be $ 1.5$ fb.
Even taking into account the selection cuts for the decay products of the
top quark, we should still have enough signal events per LHC year.

\section{Conclusions}
We have performed a comprehensive study on the decays and production
of the pseudoscalar boson $\eta$ of the simplest little Higgs model
at the LHC.  We focus on the mass range of $m_\eta < 2 m_b$ such
that the dominant decay mode is $\eta \to \tau^+ \tau^-$, followed by
$gg$ and $c\bar c$.  The decay branching ratio into $\gamma\gamma$ is
only of the order of $10^{-4}$.

The dominant production channel is gluon fusion ($gg\to \eta$),
followed by $b\bar b$ fusion ($b\bar b \to \eta$).  However, the
sole $\eta \to \tau^+ \tau^-$ in the final state may be buried
under the Drell-Yan background in this invariant mass region.
We have therefore focused on the associated production channels of
$pp \to h \eta$ and $pp \to t\bar t \eta$.  We have shown that
$t\bar t \eta$ production is in fact large enough to give a sizable
number of events while suppressing the backgrounds, the majority of
which comes from $t\bar t c \bar c$.  On the other hand,
$h \eta \to b \bar b \tau^+ \tau^-$ suffers severely from the $b\bar b
c \bar c$ background. Unless experiments can achieve a very high
rejection factor for charm quark, this channel remains
pessimistic.

\appendix
\section{Helicity amplitudes for $q\bar q \to t \bar t \eta$ and
 $gg \to t \bar t \eta$}
\begin{figure}[th!]
\centering
\includegraphics[height=8in]{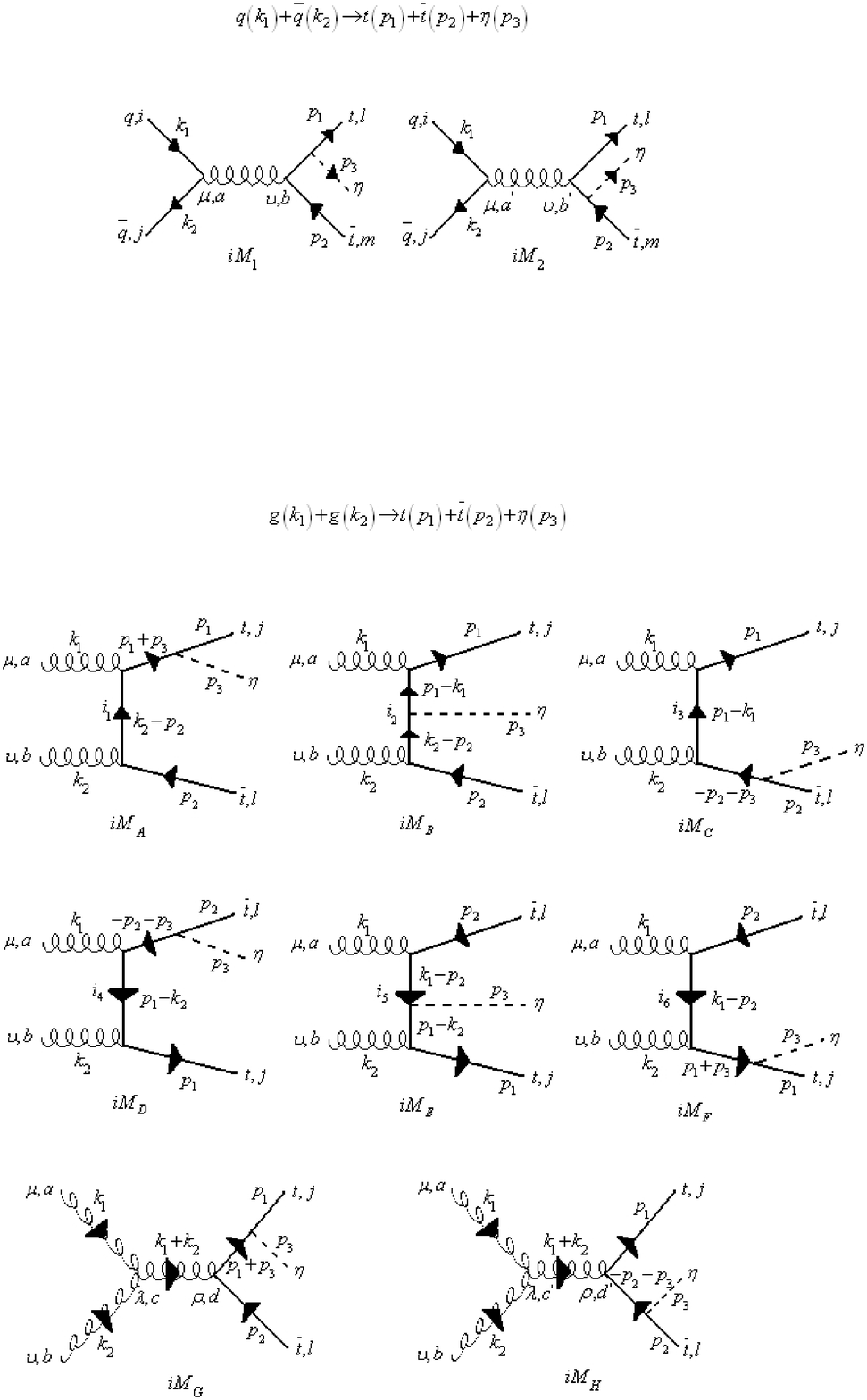}
\caption{\small \label{app-fig1}
Feynman diagrams for $q \bar q \to t \bar t \eta$ and
$g g   \to t \bar t \eta$}
\end{figure}

For the process of
\[
  q(k_1, i) + \bar q (k_2, j) \to t (p_1,l) + \bar t (p_2, m) +
\eta (p_3) \;,
\]
there are two contributing Feynman diagrams, as shown in Fig. \ref{app-fig1}.
Here the 4-momenta and the color indices ($i,j,l,m$) are given in
parenthesis.
The amplitudes are
\begin{eqnarray}
i M_1 &=& \frac{i g_{\eta t t} g_s^2 T^a_{ji} T^a_{lm}}{ (k_1 + k_2)^2
   \left( (p_1 + p_3)^2 - m_t^2 \right ) }\;
  \bar u(p_1)  \gamma^5  \not\!{p}_3 \gamma^\mu \,v(p_2) \;
  \bar v(k_2)  \gamma_\mu  u(k_1) \,,\nonumber \\
i M_2 &=& - \frac{i g_{\eta t t} g_s^2 T^a_{ji} T^a_{lm}}{ (k_1 + k_2)^2
   \left( (p_2 + p_3)^2 - m_t^2 \right ) }\;
  \bar u(p_1)  \gamma^\mu  \not\!{p}_3 \gamma^5 v(p_2) \;
  \bar v(k_2)  \gamma_\mu u(k_1)\,,  \nonumber
\end{eqnarray}
where $g_{\eta t t} = \sqrt{2} \cot 2\beta (m_t/f)$ and the interaction
Lagrangian is ${\cal L} = -i g_{\eta tt} \eta \, \bar f \gamma^5 f$.

There are eight Feynman diagrams, as depicted in Fig.\,\ref{app-fig1},
contributing to the subprocess
\[
 g (k_1,a )+ g(k_2, b) \to t (p_1, j) + \bar t(p_2,i) + \eta (p_3)
\]
where $a,b,i,j$ are color indices.
The $t$-channel-like helicity amplitudes are
\begin{eqnarray}
i M_A &=& \frac{i A}{ (2 p_1 \cdot p_3 + m^2_\eta) ( 2 k_2 \cdot p_2) } \;
 \bar u(p_1)  \gamma^5 \not\!{p}_3  \gamma^\mu  \left(
   \not\!{k}_2 - \not\!{p}_2 + m_t \right )\, \gamma^\nu \, v(p_2) \;
  \epsilon_\mu (k_1)\, \epsilon_\nu(k_2), \nonumber \\
i M_B &=& \frac{- i A}{ 4 (k_1 \cdot p_1 ) ( k_2 \cdot p_2) } \;
 \bar u(p_1) \gamma^5  \gamma^\mu
\left( \not\!{p}_1 - \not\!{k}_1 - m_t \right )
\left( \not\!{k}_2 - \not\!{p}_2 + m_t \right )
\gamma^\nu \, v(p_2) \;
  \epsilon_\mu (k_1)\, \epsilon_\nu(k_2),  \nonumber \\
i M_C &=& \frac{-i A}{ (2 p_2 \cdot p_3 + m^2_\eta) ( 2 k_1 \cdot p_1) } \;
 \bar u(p_1) \gamma^5 \gamma^\mu  \left(
   \not\!{p}_1 - \not\!{k}_1 - m_t \right ) \gamma^\nu
 \not\!{p}_3 v(p_2) \;
  \epsilon_\mu (k_1) \epsilon_\nu(k_2).
\eea
The $u$-channel-like ones are
\bea
i M_D &=& \frac{-i B}{ (2 p_2 \cdot p_3 + m^2_\eta) ( 2 k_2 \cdot p_1) } \;
 \bar u(p_1)  \gamma^5\gamma^\nu  \left(
   \not\!{p}_1 - \not\!{k}_2 - m_t \right ) \gamma^\mu
 \not\!{p}_3  v(p_2) \;
  \epsilon_\mu (k_1) \epsilon_\nu(k_2), \nonumber \\
i M_E &=& \frac{- i B}{ 4 (k_1 \cdot p_2 ) ( k_2 \cdot p_1) } \;
 \bar u(p_1) \gamma^5  \gamma^\nu
\left( \not\!{p}_1 - \not\!{k}_2 - m_t \right )
\left( \not\!{k}_1 - \not\!{p}_2 + m_t \right )
\gamma^\mu  v(p_2) \;
  \epsilon_\mu (k_1) \epsilon_\nu(k_2),  \nonumber \\
i M_F &=& \frac{i B}{ (2 p_1 \cdot p_3 + m^2_\eta) ( 2 k_1 \cdot p_2) } \;
 \bar u(p_1)  \gamma^5 \not\!{p}_3  \gamma^\nu \left(
   \not\!{k}_1 - \not\!{p}_2 + m_t \right ) \gamma^\mu v(p_2) \;
  \epsilon_\mu (k_1) \epsilon_\nu(k_2).
  \eea
The $s$-channel-line ones are
\bea
i M_G &=& \frac{E}{ (2 p_1 \cdot p_3 + m^2_\eta) ( 2 k_1 \cdot k_2) } \;
 \bar u(p_1)  \gamma^5 \not\!{p}_3  \left(
   (\not\!{k}_1 - \not\!{k}_2) g^{\mu\nu} + 2 k_2^\mu \gamma^\nu
    - 2 k_1^\nu \gamma^\mu \right )  v(p_2) \;
  \epsilon_\mu (k_1) \epsilon_\nu(k_2), \nonumber \\
i M_H &=& \frac{- E}{ (2 p_2 \cdot p_3 + m^2_\eta) ( 2 k_1 \cdot k_2) } \;
 \bar u(p_1)  \gamma^5  \left(
   (\not\!{k}_1 - \not\!{k}_2) g^{\mu\nu} + 2 k_2^\mu \gamma^\nu
    - 2 k_1^\nu \gamma^\mu \right )  \not\!{p}_3 \,v(p_2) \;
  \epsilon_\mu (k_1) \epsilon_\nu(k_2)  , \nonumber
\end{eqnarray}
where the constant factors $A,B,E$ are given by
\[
A \equiv g_{\eta t t} g_s^2 (T^a T^b)_{jl},\qquad
B \equiv g_{\eta t t} g_s^2 (T^b T^a)_{jl},\qquad
E \equiv g_{\eta t t} g_s^2  f^{abc}T^c_{jl} \;.
\]

\section{Decay of $Z'$}
The heavy gauge boson $Z'$ decays into the SM fermions ($Z' \to
f\bar{f}$) and into the Higgs and $\eta$ bosons ($Z'\to \eta H$).  We
neglect the decay of $Z' \to W^+ W^-$ since the vertex is suppressed
by $v^2/f^2$, without $\tbt$ enhancement.  In spite of non-suppressed
$Z'$-$X^+$-$X^-$ and $Z'$-$Y^0$-$\bar{Y}^0$ couplings, the mass
relation of $M_{X,Y} \approx 0.82 M_{Z'}$ does not allow the decay of
$Z' \to X^+ X^-, Y^0 \bar{Y}^0$.  Decay rates are
\bea
\Gm (Z' \to
f\bar{f}) &=& \frac{N_C}{24\pi}g_Z^2 \left[(g^f_{2R})^2 + (g^f_{2L})^2
\right]\, M_{Z'}, \\ \no \Gm (Z' \to \eta H ) &=& \frac{c_2^2}{192
  \pi} \lm^{3/2} M_{Z'},
\eea
where $\lm = 1 + (m_\eta/M_{Z'})^4
+(m_H/M_{Z'})^4 -2 (m_\eta/M_{Z'})^2 -2 (m_H/M_{Z'}) -2
(m_\eta/M_{Z'})^2(m_H/M_{Z'}) ^2$.  We refer the expression of $c_2$
and $g^f_{2R,2L}$'s for $f=u,d,c,s,b$ to Eq.\,(\ref{eq:gRL}).  Other
couplings are
\bea
g^t_{2R} &=& \frac{2}{3} \frac{x_W}{\sqrt{3-4
    x_W}}, \quad g^t_{2L} =\frac{1}{\sqrt{3-4 x_W}} \left( \frac{1}{2}
  - \frac{1}{3}x_W \right), \\ \no g^e_{2R} &=&- \frac{x_W}{\sqrt{3-4
    x_W}}, \quad g^e_{2L} =\frac{1}{\sqrt{3-4 x_W}} \left( \frac{1}{2}
  - x_W \right),\\ \no g^\nu_{2R} &=&0, \quad g^\nu_{2L}
=\frac{1}{\sqrt{3-4 x_W}} \left( \frac{1}{2} - x_W \right),
\eea

\acknowledgments
The work of JS is supported by the Korea Research Foundation Grant (KRF-2005-070-c00030).
The work of KC, PT, and QSY is supported by the NSC
under grant no. NSC 96-2628-M-007-002-MY3 and the NCTS.

\end{document}